\documentclass[dvipsnames,twocolumn,twocolappendix]{aastex63}
\usepackage{fp}
\usepackage{amsmath}

\newcommand{\spliterr}[3]{ $ #1~ \pm$ #2 \(\pm\) #3}
\newcommand{\hounits}{km s\(^{-1}\) Mpc\(^{-1}\)}
\newcommand{\LCDM}{\(\Lambda\)CDM}

\newcommand{\mtsubscript}{$m^{\mathrm{TRGB}}_{\mathrm{814}}$}

\newcommand{\mtrgbGALONE}{26.68}
\newcommand{\mtrgbGALTWO}{27.32}

\newcommand{\edgeSTATerrGALONE}{0.023}
\newcommand{\edgeSTATerrGALTWO}{0.033}

\newcommand{\edgeSYSerrGALONE}{0.029}
\newcommand{\edgeSYSerrGALTWO}{0.010}

\newcommand{\trgblum}{-4.054}
\newcommand{\trgblumSTATerr}{0.022} 
\newcommand{\trgblumSYSerr}{0.039}

\newcommand{\ACSVextinctionGALONE}{0.417} 
\newcommand{\ACSVextinctionerrGALONE}{0.045} 

\newcommand{\ACSVextinctionGALTWO}{0.028} 
\newcommand{\ACSVextinctionerrGALTWO}{0.014} 

\newcommand{\ACSIextinctionGALONE}{0.257} 
\newcommand{\ACSIextinctionerrGALONE}{0.035} 

\newcommand{\ACSIextinctionGALTWO}{0.018} 
\newcommand{\ACSIextinctionerrGALTWO}{0.009} 
\newcommand{\ZPerr}{0.02}

\newcommand{\EEerr}{0.02}

\newcommand{\ApcorerrGALONE}{0.01}
\newcommand{\ApcorerrGALTWO}{0.02}
\FPeval{\mtrgbGALONEcor}{\mtrgbGALONE+\edgeSYSerrGALONE}
\FPeval{\mtrgbGALTWOcor}{\mtrgbGALTWO+\edgeSYSerrGALTWO}

\FPeval{\distmodGALONE}{\mtrgbGALONE-\ACSIextinctionGALONE-\trgblum}
\FPeval{\distmodGALTWO}{\mtrgbGALTWO-\ACSIextinctionGALTWO-\trgblum}

\FPeval{\distGALONE}{ 10^( (\distmodGALONE) /5)/100000 }
\FPeval{\distGALTWO}{ 10^( (\distmodGALTWO) /5)/100000 } 

\FPeval{\measSTATerrGALONE}{ (\edgeSTATerrGALONE^2)^0.5 }
\FPeval{\measSTATerrGALTWO}{ (\edgeSTATerrGALTWO^2)^0.5 }

\FPeval{\calibSYSerrGALONE}{ (\ZPerr^2+\EEerr^2+\ApcorerrGALONE^2)^0.5 }
\FPeval{\calibSYSerrGALTWO}{ (\ZPerr^2+\EEerr^2+\ApcorerrGALTWO^2)^0.5 }

\FPeval{\measSYSerrGALONE}{ (\edgeSYSerrGALONE^2+\calibSYSerrGALONE^2)^0.5 }
\FPeval{\measSYSerrGALTWO}{ (\edgeSYSerrGALTWO^2+\calibSYSerrGALTWO^2)^0.5 }

\FPeval{\totSTATerrGALONE}{(\trgblumSTATerr^2+\measSTATerrGALONE^2)^0.5  }
\FPeval{\totSYSerrGALONE}{(\trgblumSYSerr^2+\measSYSerrGALONE^2+\ACSIextinctionerrGALONE^2)^0.5  }
\FPeval{\toterrGALONE}{ (\totSTATerrGALONE^2+\totSYSerrGALONE^2)^0.5 }

\FPeval{\totSTATerrGALTWO}{(\trgblumSTATerr^2+\measSTATerrGALTWO^2)^0.5  }
\FPeval{\totSYSerrGALTWO}{(\trgblumSYSerr^2+\measSYSerrGALTWO^2+(\ACSIextinctionerrGALTWO)^2)^0.5  }
\FPeval{\toterrGALTWO}{ (\totSTATerrGALTWO^2+\totSYSerrGALTWO^2)^0.5 }

\FPeval{\distGALONEupperSTAT}{ 10^( (\distmodGALONE+\totSTATerrGALONE) /5)/100000 }
\FPeval{\distGALONElowerSTAT}{ 10^( (\distmodGALONE-\totSTATerrGALONE) /5)/100000 }
\FPeval{\distSTATerrGALONE}{ 0.5*(\distGALONEupperSTAT - \distGALONElowerSTAT) }

\FPeval{\distGALTWOupperSTAT}{ 10^( (\distmodGALTWO+\totSTATerrGALTWO) /5)/100000 }
\FPeval{\distGALTWOlowerSTAT}{ 10^( (\distmodGALTWO-\totSTATerrGALTWO) /5)/100000 }

\FPeval{\distSTATerrGALTWO}{ 0.5*(\distGALTWOupperSTAT - \distGALTWOlowerSTAT) }

\FPeval{\distGALONEupperSYS}{ 10^( (\distmodGALONE+\totSYSerrGALONE) /5)/100000 }
\FPeval{\distGALONElowerSYS}{ 10^( (\distmodGALONE-\totSYSerrGALONE) /5)/100000 }
\FPeval{\distSYSerrGALONE}{ 0.5*(\distGALONEupperSYS - \distGALONElowerSYS) }

\FPeval{\distGALTWOupperSYS}{ 10^( (\distmodGALTWO+\totSYSerrGALTWO) /5)/100000 }
\FPeval{\distGALTWOlowerSYS}{ 10^( (\distmodGALTWO-\totSYSerrGALTWO) /5)/100000 }
\FPeval{\distSYSerrGALTWO}{ 0.5*(\distGALTWOupperSYS - \distGALTWOlowerSYS) }

\FPeval{\FOURTEENdiffTHIRTEEN}{ \distmodGALTWO - 31.46 }
\FPeval{\FOURTEENdiffTHIRTEENerr}{ (\measSTATerrGALTWO^2 + 0.05^2)^(1/2)}

\FPeval{\mtrgbGALONEround}{round(\mtrgbGALONE,2)}
\FPeval{\mtrgbGALTWOround}{round(\mtrgbGALTWO,2)}

\FPeval{\mtrgbGALONEcorround}{round(\mtrgbGALONEcor,2)}
\FPeval{\mtrgbGALTWOcorround}{round(\mtrgbGALTWOcor,2)}

\FPeval{\measSTATerrGALONEround}{round(\measSTATerrGALONE,2)}
\FPeval{\measSTATerrGALTWOround}{round(\measSTATerrGALTWO,2)}

\FPeval{\calibSYSerrGALONEround}{round(\calibSYSerrGALONE,2)}
\FPeval{\calibSYSerrGALTWOround}{round(\calibSYSerrGALTWO,2)}

\FPeval{\measSYSerrGALONEround}{round(\measSYSerrGALONE,2)}
\FPeval{\measSYSerrGALTWOround}{round(\measSYSerrGALTWO,2)}

\FPeval{\totSTATerrGALONEround}{round(\totSTATerrGALONE,2)}
\FPeval{\totSYSerrGALONEround}{round(\totSYSerrGALONE,2)}
\FPeval{\toterrGALONEround}{round(\toterrGALONE,3)}

\FPeval{\totSTATerrGALTWOround}{round(\totSTATerrGALTWO,2)}
\FPeval{\totSYSerrGALTWOround}{round(\totSYSerrGALTWO,2)}
\FPeval{\toterrGALTWOround}{round(\toterrGALTWO,3)}

\FPeval{\ACSVextinctionGALONEround}{round(\ACSVextinctionGALONE,2)} 
\FPeval{\ACSVextinctionGALTWOround}{round(\ACSVextinctionGALTWO,2)} 
\FPeval{\ACSIextinctionGALONEround}{round(\ACSIextinctionGALONE,2)} 
\FPeval{\ACSIextinctionGALTWOround}{round(\ACSIextinctionGALTWO,2)} 

\FPeval{\ACSVextinctionerrGALONEround}{round(\ACSVextinctionerrGALONE,2)} 
\FPeval{\ACSVextinctionerrGALTWOround}{round(\ACSVextinctionerrGALTWO,2)} 
\FPeval{\ACSIextinctionerrGALONEround}{round(\ACSIextinctionerrGALONE,2)} 
\FPeval{\ACSIextinctionerrGALTWOround}{round(\ACSIextinctionerrGALTWO,2)} 

\FPeval{\distmodGALONEround}{round(\distmodGALONE,2)}
\FPeval{\distmodGALTWOround}{round(\distmodGALTWO,2)}

\FPeval{\distGALONEround}{round(\distGALONE,1)}
\FPeval{\distGALTWOround}{round(\distGALTWO,1)}

\FPeval{\distSTATerrGALONEround}{round(\distSTATerrGALONE,1)}
\FPeval{\distSTATerrGALTWOround}{round(\distSTATerrGALTWO,1)}
\FPeval{\distSYSerrGALONEround}{round(\distSYSerrGALONE,1)}
\FPeval{\distSYSerrGALTWOround}{round(\distSYSerrGALTWO,1)}

\FPeval{\FOURTEENdiffTHIRTEENround}{round(\FOURTEENdiffTHIRTEEN,3)}
\FPeval{\FOURTEENdiffTHIRTEENerrround}{round(\FOURTEENdiffTHIRTEENerr,3)}

\received{}
\revised{}
\accepted{}

\submitjournal{ApJ}

\shorttitle{TRGB Distances to NGC~5643 and NGC~1404}
\shortauthors{Hoyt et al.}

\graphicspath{ {./figs_use/} }
\begin{document}

\title{The Carnegie Chicago Hubble Program X: Tip of the Red Giant Branch Distances to NGC~5643 and NGC~1404}

\correspondingauthor{Taylor Hoyt}
\email{tjhoyt@uchicago.edu, taylorjhoyt@gmail.com}

\author[0000-0001-9664-0560]{Taylor J. Hoyt}
\affiliation{Department of Astronomy \& Astrophysics, University of Chicago, 5640 South Ellis Avenue, Chicago, IL 60637, USA}
\affiliation{The Observatories of the Carnegie Institution for Science, 813 Santa Barbara Street, Pasadena, CA 91101, USA}

\author[0000-0002-1691-8217]{Rachael L. Beaton}
\affiliation{Department of Astrophysical Sciences, Princeton University, 4 Ivy Lane, Princeton, NJ 08544, USA}
\altaffiliation{Carnegie-Princeton Fellow}
\altaffiliation{Hubble Fellow}
\affiliation{The Observatories of the Carnegie Institution for Science, 813 Santa Barbara Street, Pasadena, CA 91101, USA}

\author[0000-0003-3431-9135]{Wendy L. Freedman}
\affiliation{Department of Astronomy \& Astrophysics, University of Chicago, 5640 South Ellis Avenue, Chicago, IL 60637, USA}

\author[0000-0002-2502-0070]{In Sung Jang}
\affiliation{Department of Astronomy \& Astrophysics, University of Chicago, 5640 South Ellis Avenue, Chicago, IL 60637, USA}

\author[0000-0003-2713-6744]{Myung~Gyoon~Lee} \affiliation{Department of Physics \& Astronomy, Seoul National University, Gwanak-gu, Seoul 151-742, Republic of Korea}

\author[0000-0002-1576-1676]{Barry F. Madore}
\affiliation{The Observatories of the Carnegie Institution for Science, 813 Santa Barbara Street, Pasadena, CA 91101, USA}
\affiliation{Department of Astronomy \& Astrophysics, University of Chicago, 5640 South Ellis Avenue, Chicago, IL 60637, USA}

\author{Andrew J. Monson}
\affiliation{Department of Astronomy \& Astrophysics, The Pennsylvania State University, 525 Davey Lab, University Park, PA 16802, USA}

\author[0000-0002-8894-836X]{Jillian~R.~Neeley} 
\affiliation{Department of Physics, Florida Atlantic University, 777 Glades Road, Boca Raton, FL 33431, USA}

\author[0000-0002-5807-5078]{Jeffrey A. Rich}
\affiliation{The Observatories of the Carnegie Institution for Science, 813 Santa Barbara Street, Pasadena, CA 91101, USA}

\author[0000-0002-1143-5515]{Mark Seibert}
\affiliation{The Observatories of the Carnegie Institution for Science, 813 Santa Barbara Street, Pasadena, CA 91101, USA}

\begin{abstract}
The primary goal of the Carnegie Chicago Hubble Program (CCHP) is to calibrate the zero-point of the Type Ia supernova (SN~Ia) Hubble Diagram through the use of Population II standard candles. So far, the CCHP has measured direct distances to 11 SNe~Ia, and here we increase that number to 15 with two new TRGB distances measured to NGC~5643 and NGC~1404, for a total of 20 SN~Ia calibrators. We present resolved, point-source photometry from new Hubble Space Telescope (HST) imaging of these two galaxies in the F814W and F606W bandpasses. From each galaxy's stellar halo, we construct an F814W-band luminosity function in which we detect an unambiguous edge feature identified as the Tip of the Red Giant Branch (TRGB). For NGC~5643, we find \( \mu_0 = \)\spliterr{\distmodGALONEround}{\totSTATerrGALONEround (stat)}{\totSYSerrGALONEround (sys)}~mag, and for NGC~1404 we find \( \mu_0 = \)\spliterr{\distmodGALTWOround}{\totSTATerrGALTWOround (stat)}{\totSYSerrGALTWOround (sys)}~mag. From a preliminary consideration of the SNe~Ia in these galaxies, we find increased confidence in the results presented in Paper VIII \citep{freedman_2019}. The high precision of our TRGB distances enables a significant measurement of the 3D displacement between the Fornax Cluster galaxies NGC~1404 and NGC~1316 (Fornax~A) equal to $1.50^{+0.25}_{-0.39}$~Mpc, which we show is in agreement with independent literature constraints.
\end{abstract}

\section{Introduction} \label{sect:intro}
Over the last decade, interest in measurements of the Hubble Constant has accelerated. According to ADS, in the years 2010, 2016, and 2020 there were, respectively, 63 (35 refereed), 108 (73 refereed), and 346 (190 refereed) article abstracts containing the phrase ``Hubble constant.'' As more measurements have accumulated in the literature and quoted uncertainties have decreased, the ``direct'' (low-redshift) and ``model-dependent'' (high-redshift) measurements of $H_0$ have diverged, clustering just above and below 70 \hounits. Strictly speaking, this no-longer nascent ``Hubble Tension'' could be the observational smoking gun for a significant departure of nature from \LCDM, i.e., the discovery of new physics beyond the standard model \citep[see, e.g.,][for an overview]{freedman_2017, riess_2020}. On the other hand, the possibility of underestimated uncertainties has yet to be ruled out definitively \citep[e.g.,][]{efstathiou_2020}. In anticipation of the degenerate nature of the $H_0$ tension (new physics or systematics), the Carnegie Chicago Hubble Program \citep[CCHP]{beaton_2016, freedman_2019} was initiated to provide an independent evaluation of the distance scale using the Tip of the Red Giant Branch (TRGB) and RR Lyrae.

The theory behind the TRGB as a standard candle has been explored thoroughly in the literature \citep[see, e.g.,][]{bellazzini_2004, salaris_2005}. In summary, the TRGB feature is a result of the ``Helium Flash,'' which is the seconds-long thermonuclear runaway event that occurs in the electron-degenerate cores of RGB stars when they reach temperatures sufficient to fuse Helium. It turns out that low mass ($ \lesssim 1 M_{\odot}$) RGB stars all reach identical levels of electron degeneracy and core mass at the time of Helium ignition, and, as a result, also reach identical surface luminosities, before then swiftly evolving off the Tip of the RGB and down to the Horizontal Branch (or the Red Clump). Ultimately, this standard luminosity at the time of the Helium Flash is mapped into the TRGB feature seen on the H-R diagram.
In observations, the TRGB can be identified in color-magnitude diagrams and stellar luminosity functions by a sharp truncation in the counts of RGB stars.

As an empirical distance indicator, the TRGB was ushered into to the modern era by some of the first observations to ever use CCDs as means to resolve the stellar populations of nearby galaxies \citep[e.g.,][]{mould_1986, mould_1988, freedman_1989, dacosta_1990}. \citet{lee_1993} formalized the method by using a Sobel edge detection kernel to locate the TRGB, with the aim of evaluating the systematic error budget in the Cepheid distance scale. \citet{sakai_2004}, \citet{mager_2008}, and \citet{mould_2009} then used the TRGB to perform additional stress tests of the Cepheid Distance Scale. \citet{karachentsev_2006}, \citet{makarov_2006}, and \citet{Rizzi_2007} developed a definitive methodology to measure and calibrate the TRGB method, particularly, but not necessarily, in the context of HST imaging. These techniques have been employed extensively by the Cosmicflows group \citep{jacobs_09,tully_09,tully_13,tully_16} to measure TRGB distances to $\sim 500$ nearby galaxies as part their program to constrain the nearby velocity field. \citet{freedman_2010}, in their review of the Hubble Constant, highlighted the TRGB method's potential to establish a fully independent Population II distance scale that is comparable in precision to the Cepheid distance scale.

The Carnegie Chicago Hubble Program (CCHP) has since used the Hubble Space Telescope (HST) to measure TRGB distances to galaxies that have hosted a SN~Ia \citep{jang_2018,hatt_2018a, hatt_2018b, hoyt_2019, beaton_2019}. These SN calibrator galaxies are all located within $\sim \! 20$~Mpc because they must be nearby enough to detect, resolve, and accurately photometer RGB stars contained in their stellar halos. Additionally, observations of RR Lyrae and TRGB stars within the Galaxy and the Local Group have been used to validate and augment our nearby calibration of the distance scale \citep{hatt_2017,rich_2018,neeley_2019}.

At the culmination of its first phase, the CCHP presented TRGB distances to ten individual SN host galaxies \citep{freedman_2019} and combined those with five others \citep{Jang_2017a,Jang_2017b} to calibrate the SNe~Ia observed by the Carnegie Supernova Project (CSP). They measured $H_0 =$\spliterr{69.8}{0.8(stat)}{1.7(sys)} \hounits. This value, based on a geometric calibration of the TRGB method via detached eclipsing binaries (DEBs) in the Large Magellanic Cloud \citep[LMC,][]{pietrzynski_2019} was updated in \citet{freedman_2020} to 69.6 \hounits. The CCHP measurement of $H_0$ is in better agreement with the Planck high-redshift value \citep[][$67.4 \pm 0.5$ \hounits; $1.1\sigma$]{planck_2018}, than with the SH0ES low-redshift value \citep[][$74.02 \pm 1.42$ \hounits; 1.9 $\sigma$]{riess_2019}. The CCHP's LMC-based determination of the TRGB $I$-band absolute magnitude appears to have been confirmed via HST observations made in the stellar halo of NGC~4258 \citep{Jang_2021}, to which an updated geometric distance has recently been measured using the maser clouds orbiting its central black hole \citep{reid_2019}.

We note that in their maser distance paper, \citet{reid_2019} also proposed an alternative calibration of the CCHP distance scale. 70\% of their adopted average TRGB magnitude \citep{Macri_2006} was determined from observations that were specifically designed to discover Cepheid variables in the star-forming disk of NGC~4258. In the same paper, \citeauthor{Macri_2006} used these Cepheids to directly determine a non-zero extinction along the line of sight. By contrast, \citeauthor{reid_2019}, in their proposed TRGB re-calibration, adopted a value of zero for the \textit{in situ} reddening to this same star-forming disk region. Their criteria for ignoring reddening effects to this disk field included its \emph{radial} separation from the galaxy's center, which is a suitable approximation for observations positioned along the minor axis of a disk (such as those designed by the CCHP), but not for those authors' preferred TRGB imaging of NGC~4258, which is aligned with the major axis of its highly inclined ($b/a > 2$) disk. We refer the reader to Section 4.2 of \citet{Jang_2021} for an extensive, multi-wavelength (from radio to FUV) examination of this disk field that \citet{reid_2019} used to calibrate the TRGB, and why an accurate TRGB magnitude cannot be measured from it.\footnote{See also Section 6 of \citet{Rizzi_2007} where the same ``Outer disk'' imaging data were used to determine a TRGB magnitude that is 0.1~mag offset from, and where the uncertainty was estimated to be at least three times larger than, that adopted by \citet{reid_2019}.}

With the TRGB zero-point calibration seemingly well understood at this time, we turn to the middle rung of the distance ladder -- a zero-point calibration of the SN~Ia luminosity-width relation -- to continue exploring the persistent discrepancies that exist between different $H_0$ measurements, as well as to increase the precision of the existing CCHP measurement. We present here two TRGB distance determinations made to the nearby galaxies NGC~5643 and NGC~1404, each of which has hosted two SNe~Ia. The TRGB distances presented here are the first published distances to these galaxies determined with a primary (resolved stars) distance indicator such as the Cepheids or the TRGB.\footnote{In the final stages of preparation of this manuscript, an (unpublished) TRGB distance to NGC~5643 using the data presented here was released on the Extragalactic Distance Database (EDD).}

NGC~5643 is a Seyfert Type II, luminous infrared galaxy viewed nearly face-on, and is located at galactic latitude $b= +15^{\circ}$. According to the NASA Extragalactic Database (NED), the only prior distances measured to NGC~5643 were made using the Tully-Fisher relation \citep{bottinell_1984,bottinell_1985,bottinell_1986,Tully_1988}. More recently, NGC~5643 has hosted a pair of SNe Ia, SN~2013aa \citep{parker_2013} and SN~2017cbv \citep{hosseinzadeh_2017, Coulter_2017, tartaglia_2017}; \citet{burns_2020} found both SNe to agree to 3\% in distance. 

NGC~1404 is a massive elliptical galaxy that belongs to the Fornax Cluster. It is host to dozens of known companion galaxies \citep{ferguson_1989,derijcke_2003,mieske_2007} and hundreds of globular clusters \citep{forbes_1998}. It has been shown with X-ray observations that NGC~1404 is actively falling into the gravitational well of the Fornax cluster \citep{machacek_2005}.
According to NED, the only prior distances to NGC~1404 come via secondary distance indicators. These include the globular cluster luminosity function \citep[GCLF,][]{Richtler_1992, Blakeslee_1996, Grillmair_1999, Ferrarese_2000, Gomez_2001, Humphrey_2009, Villegas_2010}, the planetary nebula luminosity function \citep[PNLF,][]{McMillan_1993,Ferrarese_2000,Ciardullo_1993,Ciardullo_2002}, surface brightness flucuations \citep[SBF,][]{Buzzoni_1993,Ciardullo_1993,Ferrarese_2000,liu_2002,Tonry_1991,Tonry_2001, Jensen_1998,Jensen_2001,Jensen_2003,blakeslee_2001,Blakeslee_2009,blakeslee_2010}, SNe Ia \citep{Weyant_2014, Dhawan_2016, Hoeflich_2017, gall_2018}, and the Tully-Fisher relation \citep{Theureau_2007}. NGC~1404 was host to SN~2007on and SN~2011iv which were found by \citet{gall_2018} to differ in their respective distances by 9\% in the NIR and 14\% in the optical. These SNe were already included in the \citet{freedman_2019} $H_0$ analysis, where a distance estimate was based on the average of direct TRGB distances measured to NGC~1316 and NGC~1365, both of which are also members of the Fornax Cluster. In this study, we update that distance to NGC~1404 with a \textit{direct} TRGB determination.

This paper is organized as follows: in \autoref{sect:data} we present the imaging, photometry, and color-magnitude diagrams (CMDs). Then, in \autoref{sect:trgb} we step through the determination of a TRGB distance to each galaxy. In \autoref{sect:snia_calib} we re-compute the CCHP SN~Ia calibration with the updated distance to NGC~1404. In \autoref{sect:discussion} we discuss and compare with previous distance measurements (all using secondary distance indicators). We conclude in \autoref{sect:conclusion}, and in the Appendix we present photometry for auxiliary objects of interest as well as a more comprehensive discussion on the literature distance comparisons. 

\section{Data} \label{sect:data}
\subsection{HST Observations} \label{subsect:obs}

Observations were taken with the Advanced Camera for Surveys (ACS) Wide Field Camera (WFC) on board the Hubble Space Telescope (HST). Four orbits were dedicated to the stellar halo of NGC 5643 in each of the F814W and F606W bandpasses. For NGC 1404, 17 ACS/WFC orbits in each of the F814W and F606W bands were aimed slightly NW of, but still including, the elliptical galaxy's nucleus. The observations are summarized in \autoref{tab:obs}.

\begin{figure*}
    \centering
    \includegraphics[width=0.7\textwidth]{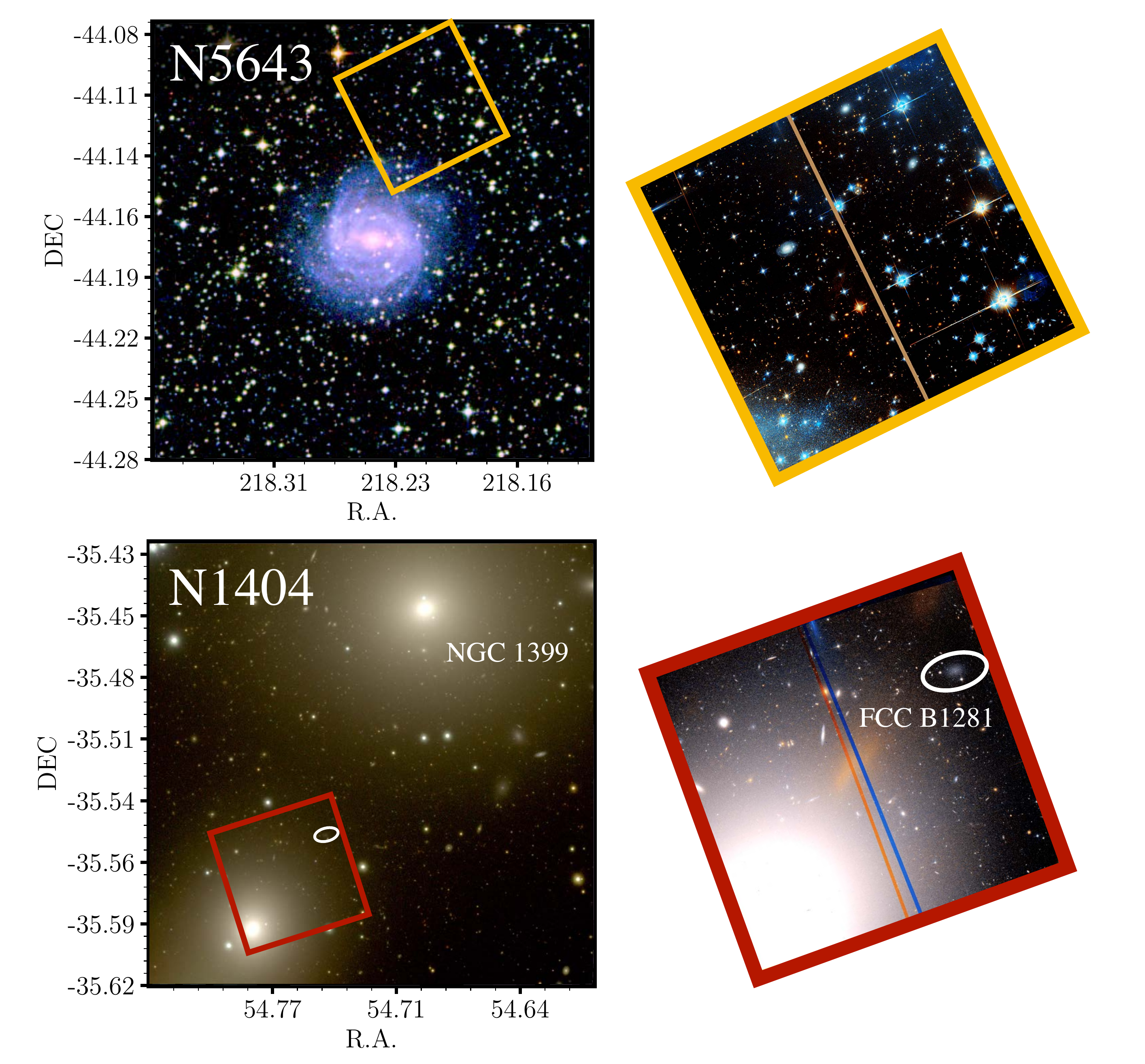}
    \caption{\textit{left}: False-color, ground-based images of NGC~5643 and NGC~1404 from the DSS and Fornax Deep Survey (FDS), respectively. Outlined are ACS imaging footprints. In the NGC~1404 image, we have labeled two additional galaxies of interest: NGC~1399 and FCC~B1281. \textit{right}: HST/ACS imaging used to determine the TRGB distance to each galaxy.
    }
    \label{fig:pointings}
\end{figure*}

In the left column of \autoref{fig:pointings}, we show the HST/ACS imaging footprints in red, overlaid on ground-based false-color images that contain each target. For NGC~5643 we show a DSS\footnote{The Digitized Sky Surveys were produced at the Space Telescope Science Institute under U.S. Government grant NAG W-2166. The images of these surveys are based on photographic data obtained using the Oschin Schmidt Telescope on Palomar Mountain and the UK Schmidt Telescope. The plates were processed into the present compressed digital form with the permission of these institutions.} 
Blue/Red/Far-red image, and for NGC~1404 we show a \textit{gri} image made from Fornax Deep Survey \citep{venhola_2018} imaging.\footnote{Based on data obtained from the ESO Science Archive Facility.}
The right column of images in \autoref{fig:pointings} are pseudo-RGB (F606W, (F606W+F814W)/2, F814W) representations of the HST/ACS imaging.

In the RGB images of NGC~5643 shown in \autoref{fig:pointings} a large number of blue foreground stars can be seen, a result of its position at a galactic latitude $b= +15^{\circ}$. We can identify disk, bar, and spiral arm structures in this late-type spiral galaxy, in addition to a well-defined transition from disk-dominated to halo-dominated components. A portion of the disk is still evident in the SE corner of the ACS imaging. In \autoref{sect:spatial}, we will describe our procedure for determining the extent to which the disk must be masked in order to measure predominantly halo RGB stars. Additionally, there is a faint blue glow visible in the NW corner of the ACS imaging that we suspect is due to scattered light in the optics or diffuse glow from a Galactic substructure (see \autoref{sect:spatial} for why we do not think it is a structure belonging to NGC~5643).

The $gri$ image that contains NGC~1404 ($V = 10.00 \pm 0.13$~mag) also reveals $10'$ to the NW the giant elliptical NGC~1399, the central bright galaxy of the Fornax Cluster\footnote{NGC~1316 is the brightest, but not located at the dynamical center of the cluster.} with $V = 9.59 \pm 0.10 $~mag \citep{dePaz_2007}. In designing the observing program, we needed to take into account the halo of NGC~1399, which \citet{iodice_2016} showed extends as far as $33'$ ($\sim \! 180$ kpc), as well as the $V = 8.09 $~mag star HD~22862 located $3'$ SE of NGC~1404 (not pictured). So we aimed HST at the NW quadrant of the main body of NGC~1404 and not in its outskirts, as would have otherwise been ideal for a TRGB measurement. As a result, unresolved light from the nucleus of NGC~1404 dominates in the innermost chip (Chip 2). The pseudo-color representation of the $HST$ imaging shown in the bottom-right panel of \autoref{fig:pointings} also reveals a diffuse, red light extending from the major axis of the galaxy nucleus. This nebulous light in the ACS imaging is most prominent where seen bridging the chip gap, but also can be seen in the NW corner. These artefacts are elongated along the same axis as the nucleus and are much redder than the unsaturated light from the galaxy, providing evidence that they are optical artefacts produced by the saturated nucleus of NGC~1404, and not astrophysical structures.

Additionally, $HST$ could not acquire guide stars for a portion of its visit to this field, which led to a small positional offset between the observations taken in each bandpass. In the bottom-right panel of \autoref{fig:pointings}, the effect can be identified by the unaligned chip gaps of two different colors. Despite this, we found no evidence of complications (e.g., blurring of the PSF) within each exposure.

\begin{deluxetable*}{lllrrll}
\tabletypesize{\normalsize} 
\tablewidth{0pt} 
\tablecaption{ACS/WFC Observation Summary\label{tab:obs}} 
\tablehead{ 
\colhead{Target Name} &
\colhead{Morph. Type\tablenotemark{a}} &
\colhead{Observation Dates} &
\colhead{F606W}&
\colhead{F814W} &
\colhead{RA\tablenotemark{b}} &
\colhead{Dec\tablenotemark{b}} \\
\colhead{} &
\colhead{} &
\colhead{} &
\colhead{($N_{orbits} \times $ sec.)}&
\colhead{($N_{orbits} \times $ sec.)} &
\colhead{(h:m:s)} &
\colhead{(d:m:s)} 
}
\startdata 
NGC 5643 & SAB(rs)c & 2019 Aug 22-23 &  $4 \times 2400$ & $4\times 2400$ & 14:32:42.340 & -44:06:53.97  \\
NGC 1404 & E1 & 2019 Aug 29, Sep 15-19 &  $17\times 2400$ &  $17\times 2400$ &  03:38:46.057 & -35:34:08.27
\enddata 
\tablenotetext{a}{\citet{deVau_1991}} 
\tablenotetext{b}{J2000 coordinates at the center of the HST pointings, referenced to the first F814W exposure in each image set.}
\end{deluxetable*}
\subsection{Photometry} \label{subsect:photometry}
Here, we briefly summarize the CCHP photometry pipeline, and we refer the reader to \citet{beaton_2019} for a detailed description. We performed point-spread function (PSF) photometry on the HST images with an automated DAOPHOT/ALLFRAME \citep{stetson_1987, stetson_1994} pipeline that utilizes TinyTim theoretical PSFs \citep{krist_2011}. Aperture corrections to a 10 pixel ($0.5''$) radius were computed and applied to each frame individually. The CCHP pipeline computes aperture corrections on a frame-by-frame basis. This is intended to improve the accuracy of our final averaged photometry by better tracking variations in the focus over time.

\begin{deluxetable*}{ccrrrrrrrrrrrr}
\tabletypesize{\normalsize} 
\tablewidth{0pt} 
\tablecaption{Corrections from PSF to Aperture magnitudes at $0.5''$ \label{tab:apcors}} 
\tablehead{ 
\multicolumn{2}{c}{Target} &
\multicolumn{6}{c}{F606W}  &
\multicolumn{6}{c}{F814W}   \\
\colhead{} &
\colhead{} &
\multicolumn{3}{c}{Chip 1} &
\multicolumn{3}{c}{Chip 2} &
\multicolumn{3}{c}{Chip 1} &
\multicolumn{3}{c}{Chip 2} \\
\colhead{} &
\colhead{$N_{frames}$} &
\colhead{ApCor} &
\colhead{$\sigma$} &
\colhead{$N$} &
\colhead{ApCor} &
\colhead{$\sigma$} &
\colhead{$N$} &
\colhead{ApCor} &
\colhead{$\sigma$} &
\colhead{$N$} &
\colhead{ApCor} &
\colhead{$\sigma$} &
\colhead{$N$} \\
\colhead{} &
\colhead{} &
\colhead{(mag)} &
\colhead{(mag)} &
\colhead{} &
\colhead{(mag)} &
\colhead{(mag)} &
\colhead{} &
\colhead{(mag)} &
\colhead{(mag)} &
\colhead{} &
\colhead{(mag)} &
\colhead{(mag)} &
\colhead{} 
}
\startdata 
NGC~5643 & 8 &  0.01 & 0.01 & 20(1) & 0.04 & 0.01 & 20(1) & $-0.08$ & 0.01 & 22(3) & $-0.03$ & 0.01 & 20(3) \\
NGC~1404 & 34 &  0.02 & 0.02 & 4(1) & \nodata & \nodata & \nodata & $-0.07$ & 0.02 & 2(1) &\nodata & \nodata & \nodata
\enddata 
\tablecomments{Corrections applied as $m_{0.5''} = m_{PSF} + \mathrm{ApCor}$. Due to a lack of isolated bright sources in NGC~1404 Chip 2, we could not directly determine its PSF-to-aperture corrections. Instead, the aperture corrections measured for Chip 1 were applied to both chips.}
\end{deluxetable*}
The initial photometry catalogs contained a total of 107009 sources in the NGC~5643 imaging, with 63448 and 181566 sources detected in Chip 1 and Chip 2 of the NGC~1404 imaging. 
To clean the catalogs, we make cuts on morphology parameters returned by DAOPHOT: the magnitude error, $\chi$, and \textit{Sharp}. To do this, the automated pipeline fit exponential curves to the distributions of these parameters as a function of both F606W and F814W magnitude, and clipped those sources that lie outside of the envelopes of these exponential distributions. See Figure 2 of \citet{beaton_2019} for an example in M101.
The cleaning process left 49104 sources in the NGC~5643 catalog, as well as 48143 and 131116 sources in the NGC~1404 Chip 1 and Chip 2 catalogs. Importantly, the photometry cleaning process had no measurable effect on the final TRGB distance determined here. This is expected because the cleaning process predominantly targets sources near the limiting magnitude of the photometry. However, because parts of our analysis depend on the photometry of these stars fainter than the TRGB magnitude (see \autoref{sect:spatial}), we choose to adopt the cleaned catalogs going forward.

We also required that a source be detected in a minimum of 6 (75\%) and 17 (50\%) frames in the NGC 5643 and NGC 1404 imaging, respectively. This requirement left 40031, 48060, and 130934 in the NGC~5643, NGC~1404-C1, and NGC~1404-C2 photometry, respectively. We note that we varied the fraction of frames in which a source must be detected between 30\% and 88\%, and observed no change to our final TRGB distances.

Finally, we made a conservative cut on the sky background determined locally to each detected source, which we use to mask extended and saturated sources that survived our source morphology cuts. We did not apply this sky background cut to the Chip 2 photometry because unresolved light from the nucleus dominates in that chip, leading to background counts that span three orders of magnitude, from the saturation limit of the chip to a few dozens of ADU. If we were to apply the same sky cut that was made on the Chip 1 catalog, only 27374 sources (20\%) would remain in the Chip 2 photometry. This final cut left 38888 and 45843 sources in the NGC~5643 and NGC~1404-C1 catalogs, respectively, while the NGC~1404-C2 catalog retained its 130934 sources.

Due to the low galactic latitude of NGC~5643, there were plenty of bright, isolated stars from which the automated pipeline could determine an aperture correction, determined there to better than $0.01$ mag. In NGC~1404 on the other hand, the algorithm to select for candidate aperture correction stars was significantly contaminated by barely-resolved globular clusters (GCs). We manually excluded these GCs from our standard star sample by inverting a color-magnitude selection (described in the Appendix). This reduced our sample of aperture correction stars for Chip 1 from twenty to two/four, depending on bandpass. In the Chip 2 photometry, we were unable to identify any isolated, bright stars suitable for determining PSF-to-aperture corrections, which is due to the high source density observed in this chip. So we instead choose to apply the Chip 1 aperture corrections to the Chip 2 photometry. Aperture corrections for the two ACS chips should be close to (though not exactly) equal, and we expect this to be the most reliable way to calibrate the Chip 2 photometry. Note that we do not use the Chip 2 photometry at all in our determination of the TRGB distance to NGC~1404. In \autoref{tab:apcors}, we present the mean and standard deviations of the aperture corrections determined for each chip, band and target.

We then calibrate the photometry to the ST VEGAMAG system by applying encircled energy corrections to an infinite aperture \citep[presented in][]{bohlin_2016} and infinite aperture photometric zero-points provided by STScI.\footnote{\url{https://acszeropoints.stsci.edu/}}

\subsection{Color Magnitude Diagrams}

\begin{figure*}
    \centering
    \includegraphics[width=0.85\textwidth]{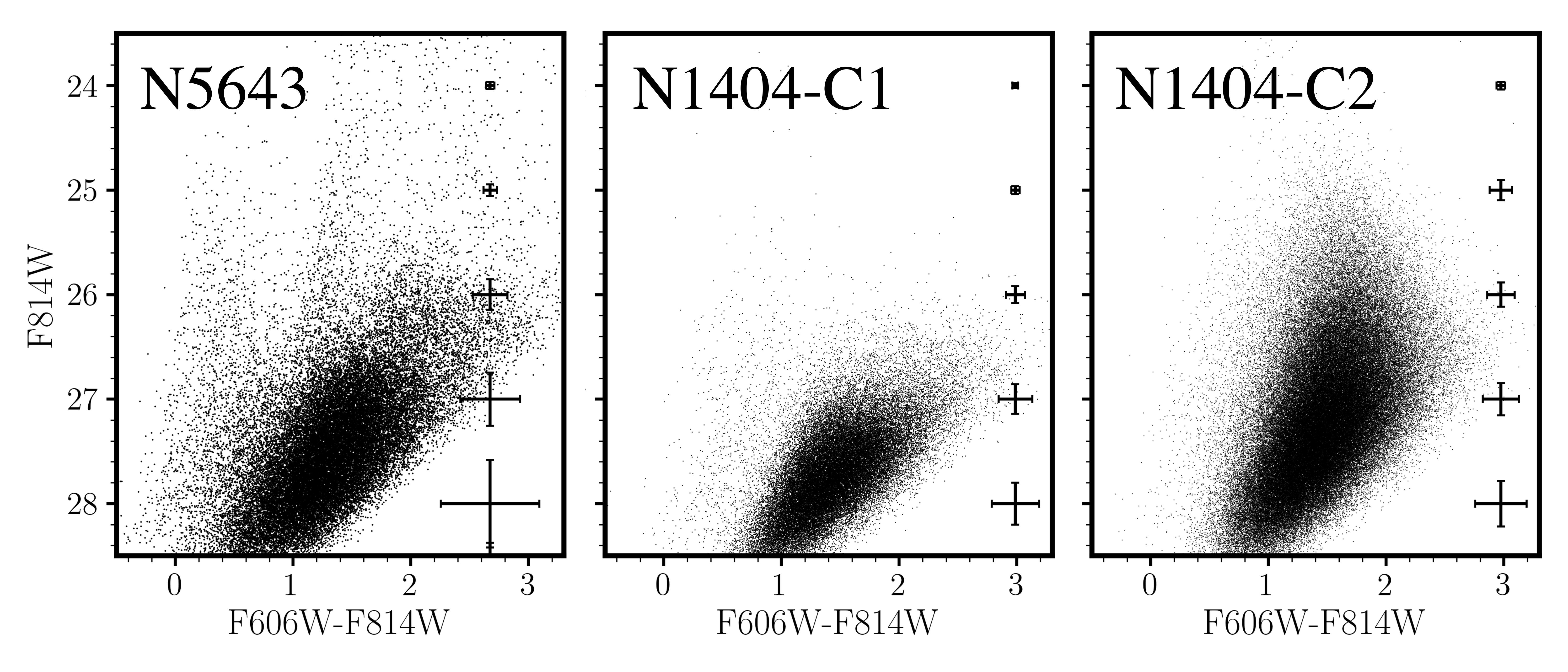}
    \caption{Color-magnitude Diagrams for NGC 5643 (left), and NGC 1404 Chip 1 (middle), Chip 2 (right). Average error bars in 1 mag intervals for color and magnitude are plotted on the right hand side of each CMD. Note that the markers are three times larger (in 1D) in the N5643 CMD than in the N1404 CMDs.
    }
    \label{fig:cmds}
\end{figure*}

In \autoref{fig:cmds} we present cleaned F606W and F814W color-magnitude diagrams (CMDs) of photometry for NGC~5643 and NGC~1404, where we have separated the latter by chip, with the corresponding labels N1404-C1 and N1404-C2.

For NGC~5643, the CMDs indicate that young, blue stars (F606W$-$F814W$\simeq 0.2 $ mag), as well as red supergiants (F606W$-$F814W$\simeq 1.0 $ mag), are present in the imaging. In \autoref{sect:spatial}, we confirm that these sources, typically classified as belonging to young stellar populations, are associated with the disk feature visible in the SE corner of the HST imaging.

In the N1404-C2 CMD, thousands of sources can be seen forming a vertical plume centered around (F606W$-$F814W) $ \simeq 1.5$ mag that extends several magnitudes above the TRGB magnitude we would expect based on the distance to the Fornax Cluster (F814W $\sim 27.5$~mag). Because this feature is not present in the Chip 1 photometry at all, has an average color equal to a typical RGB star, and appears as a disjoint population, we conclude that these sources are likely source blends of fainter RGB stars.

Additionally, we identify $\sim \!\!\!250$ globular cluster candidates, selected by their photometric colors (F606W$-$F814W) $ \simeq 1.00$ mag and F606W magnitudes. They can be seen forming a narrow vertical sequence extending from F814W$=24-20$~mag, beyond the boundaries of the N1404-C1 and N1404-C2 panels in \autoref{fig:cmds}. In the Appendix, we present details of the color-magnitude selection and photometry for these GC candidates.

Also in the Appendix, we present photometry of sources belonging to FCC~B1281 (marked in \autoref{fig:pointings}) and measure a TRGB distance to this dwarf satellite. The distance we measured agrees well with our TRGB distance to NGC~1404, which provides a sanity check on the precision of our distance measurement, and confirms the membership of FCC~B1281 to NGC~1404.

\section{Measurement of TRGB Distances} \label{sect:trgb}

In this section we cover our determination of the TRGB distances to NGC~5643 and NGC~1404. First, in \autoref{subsect:edge}, we introduce the edge detection methodology used to determine the TRGB magnitude. In \autoref{sect:spatial} we refine our measurement by identifying for each galaxy a sample of old, metal-poor RGB stars. Then we use artificial star experiments to estimate the uncertainties associated with the edge detection in \autoref{subsect:artstar}. Finally, in \autoref{subsect:dists}, we present true TRGB distances measured to each galaxy.

\subsection{Edge Detection} \label{subsect:edge}
In \citet{hatt_2017} we introduced a method of standardizing the representation of the RGB luminosity function (LF) for the purposes of detecting the TRGB. We first constructed the LF with a bin size of 0.01 mag, which is much smaller than the typical photometric error ($\sim\!\!\!0.10$~mag) of a TRGB star in our data. We then smoothed the LF by fitting a second order polynomial about each bin's center. The fit domain included the entirety of the dataset (i.e., a non-truncated window), and each datapoint in the fit was weighted by a Gaussian \citep[GLOESS,][]{persson_2004} centered on each interpolated bin. The width of the Gaussian weighting window was then set equal to the typical photometric error of a single source at the magnitude of the expected TRGB, and then optimized using artificial star simulations. Assuming that the TRGB feature is a step function in the limit of infinite signal to noise, this choice is expected to minimally distort the location of the TRGB edge, while sufficiently suppressing higher frequency noise.

\begin{figure*}
    \centering
    \includegraphics[width=0.9\textwidth]{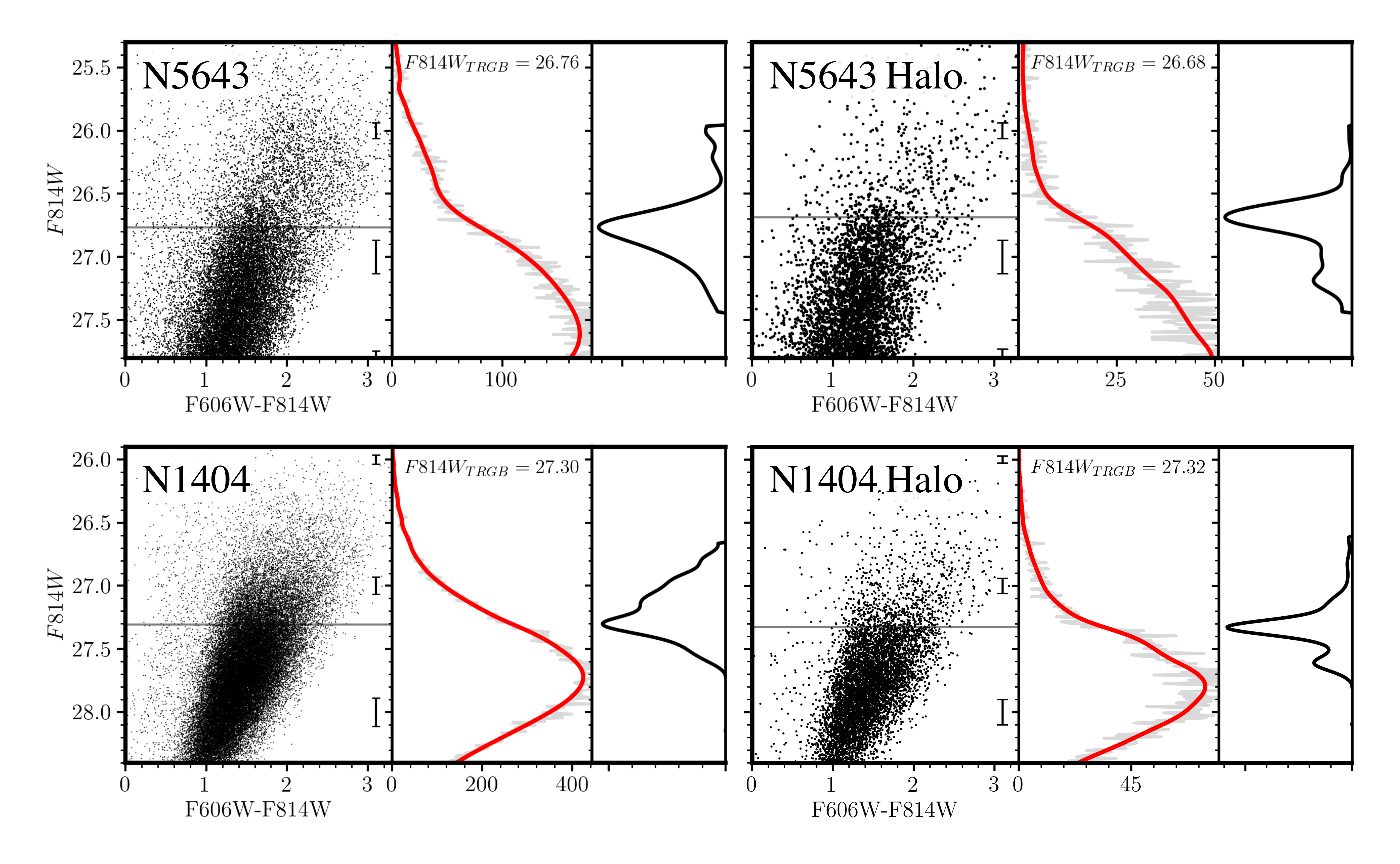}
    \caption{Edge detection measurement of the TRGB. In all four quadrants, plotted in the left panel is a color-magnitude diagram, then in the middle panel is the marginalized, smoothed luminosity function in red (with unsmoothed in gray), and in the right panel the edge response function, the peak of which is taken to be the observed TRGB magnitude. Two sets of detections are shown for each galaxy: one before any kind of spatial filtering, and another after selection for halo stars (see \autoref{sect:spatial} for details).}
    \label{fig:detections}
\end{figure*}

Next, to measure the location of the TRGB magnitude we pass a Sobel \{-1, 0, +1\} difference filter over the smoothed LF. Each point $\delta_i(m_i)$ in the resultant discrete first derivative is then multiplied by a Poisson weighting scheme $(N_{i+1} - N_{i-1}) / \sqrt(N_{i-1} + N_{i+1})$ associated with the corresponding bin in the LF. The resulting TRGB response function is displayed in the right-most column of \autoref{fig:detections}.

In this paper we make one minor change to the CCHP edge detection methodology. We use a simple Gaussian kernel to smooth the LF instead of the GLOESS interpolation method described above. Because the LFs for both galaxies are well populated (with no gaps), we observed no difference between a simple kernel smoother vs. using the GLOESS interpolation scheme; both gave identical results when detecting the TRGB edge.

In the left column of \autoref{fig:detections}, we show the results of the TRGB edge detection for the full, cleaned catalogs. The luminosity function for NGC~5643 contains many AGB stars, and the edge response for NGC~1404 is broad with a long tail to brighter magnitudes. We suspect this tail in the edge response is due to crowded RGB stars migrating above the true TRGB, which are producing false edges in the LF. In \autoref{sect:spatial}, we will show that both TRGB detections can be improved by selecting for stars in the outer regions of each imageset. In the right column, we show the TRGB detections for those sources contained in the outer region selections.

\subsection{Identifying the Old, Metal-poor RGB} \label{sect:spatial}
In order to accurately measure a TRGB distance, we must classify and isolate old, metal-poor stellar populations in the HST photometry. To do this, we invoke an analysis similar to that presented in \citet{hoyt_2019} and \citet{mager_2020}, in which a running cut on galactocentric radius was used to probe the effects of source location (relative to the host galaxy) on measurement of the TRGB. Here, we expand on this methodology by parametrizing the spatial dimension by semi-major axis (SMA) of an elliptical profile adopted for each galaxy, as opposed to galactocentric radius. We then divide the photometry into regions with boundaries that are defined in such a way as to produce equal signal strength in each region's TRGB edge.

The bounds of each elliptical region are iteratively optimized such that the quantity $N_{RGB} - N_{AGB} = constant$, where $N_{RGB}$ and $N_{AGB}$ are defined as the number of sources with (F606W$-$F814W) $ > 1.00 $ mag, and that lie within 1 mag fainter and brighter, respectively, of a single adopted value for the TRGB magnitude. We expect the difference between $N_{RGB}$ and $N_{AGB}$ to approximate to first-order the observed signal in the TRGB edge, since contaminant AGB stars that lie under the RGB will linearly decrease the contrast of the true TRGB discontinuity. We do not re-determine within each spatial bin the TRGB magnitude used to define this calculation. That is because the unbiased TRGB magnitude \emph{should} be a constant across all regions. Instead, in order to refine the determination of these quantities, we complete the following analysis twice, and use the results from the first iteration to update our estimate of the true TRGB magnitude in each galaxy. The figures discussed in this section were generated using the results of the second iteration. 

\begin{figure}
\centering
  \includegraphics[width=0.48\textwidth]{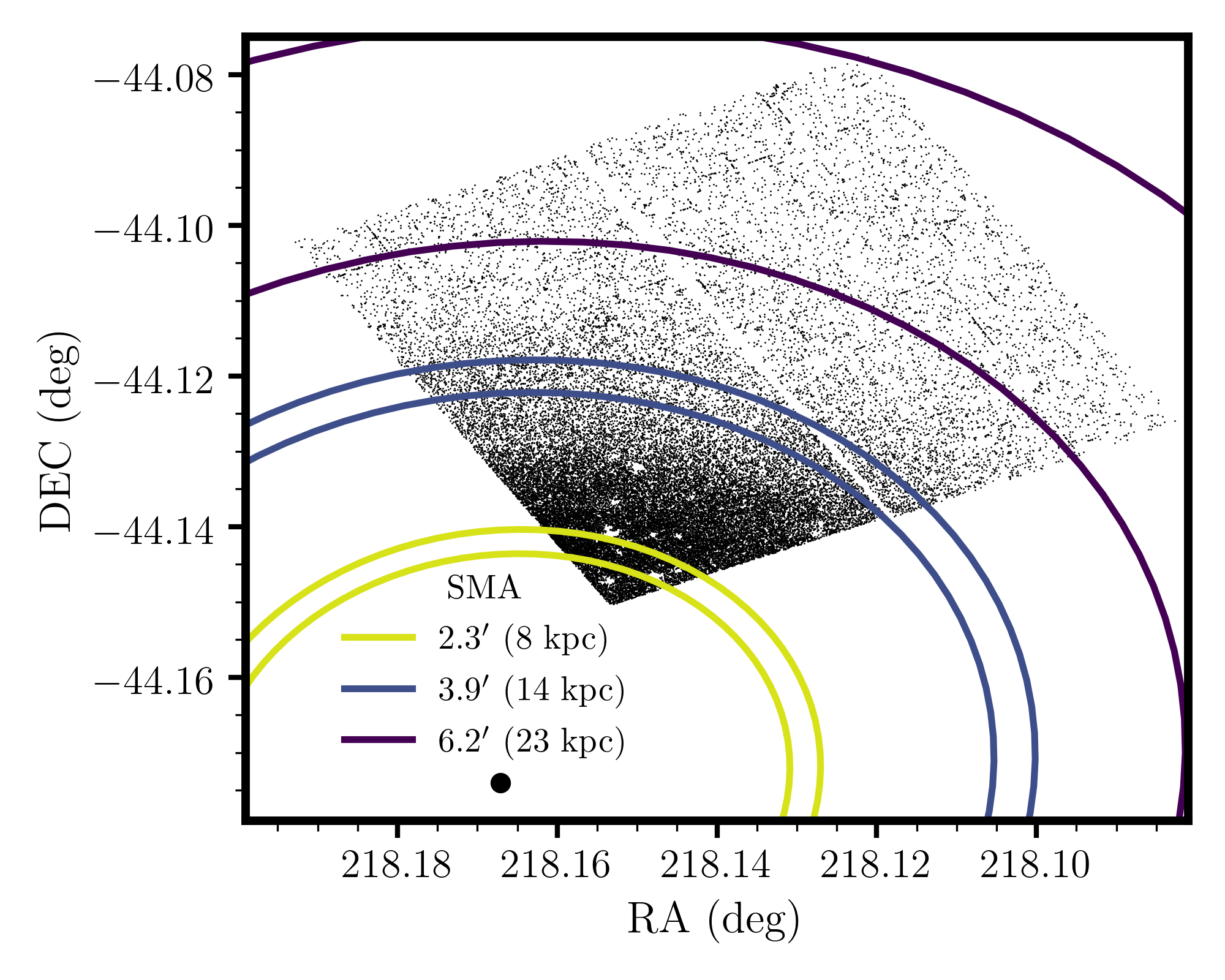}
  \includegraphics[width=0.48\textwidth]{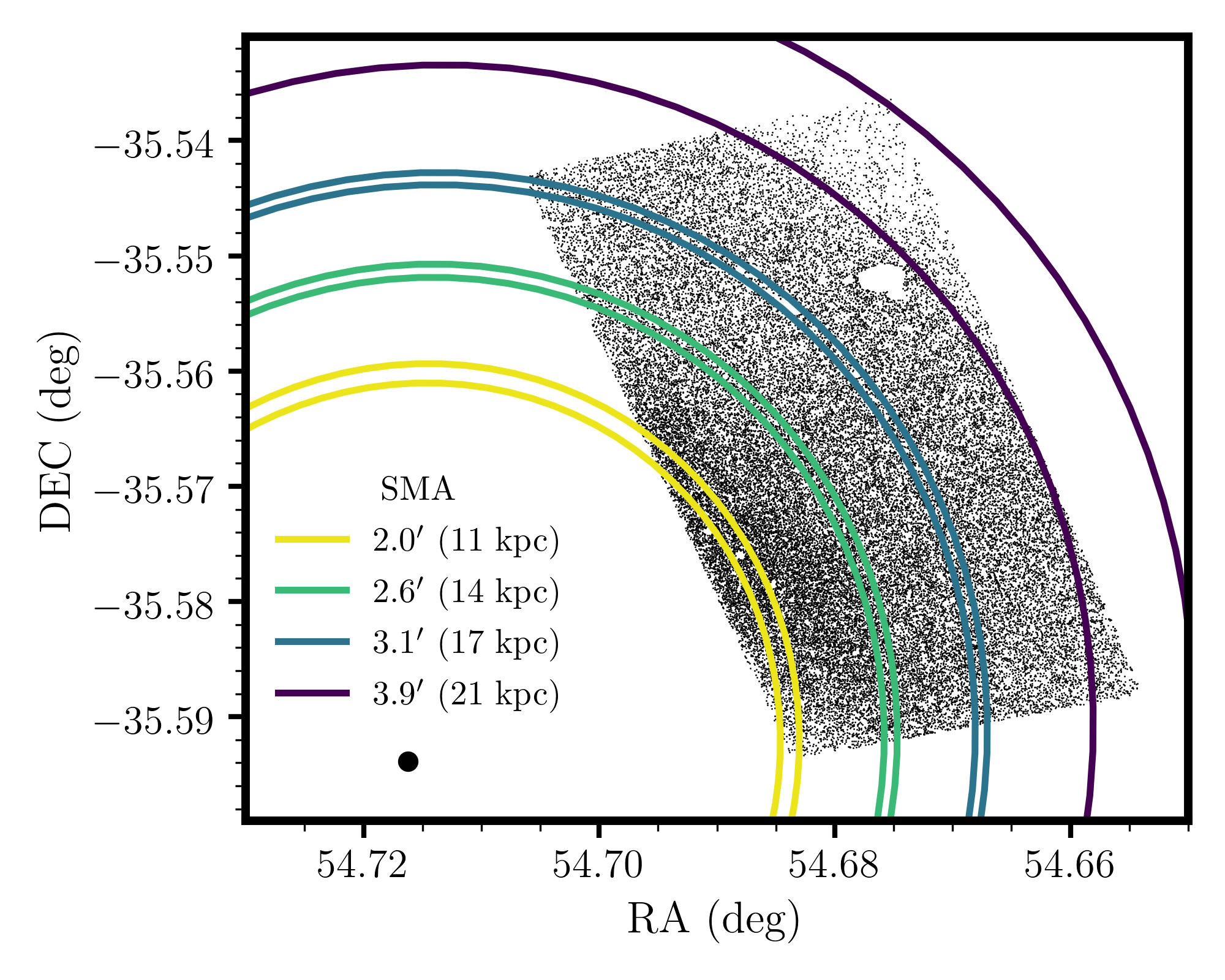}
  \caption{NGC~5643 (top), NGC~1404 (bottom). Source density maps and representative annular regions color-coded to match the color mapping used in \autoref{fig:spatial_diag}. The large black dots mark the adopted center of each galaxy's elliptical profile. All regions beyond the blue annulus in each map ($SMA = 3.9'$ and $3.1'$ for NGC~5643 and NGC~1404, respetively) are combined and used to measure the halo-based TRGB distance to each galaxy.\label{fig:spatial_regions}}
\end{figure}

In \autoref{fig:spatial_regions}, we show the results of the region determination algorithm, where a subset of the annular ellipses are plotted over a stellar source density map for each galaxy. In each panel, the large black dot marks the centroid of the elliptical profile adopted.

\begin{figure*}
  \centering
   \includegraphics[width=0.5\textwidth]{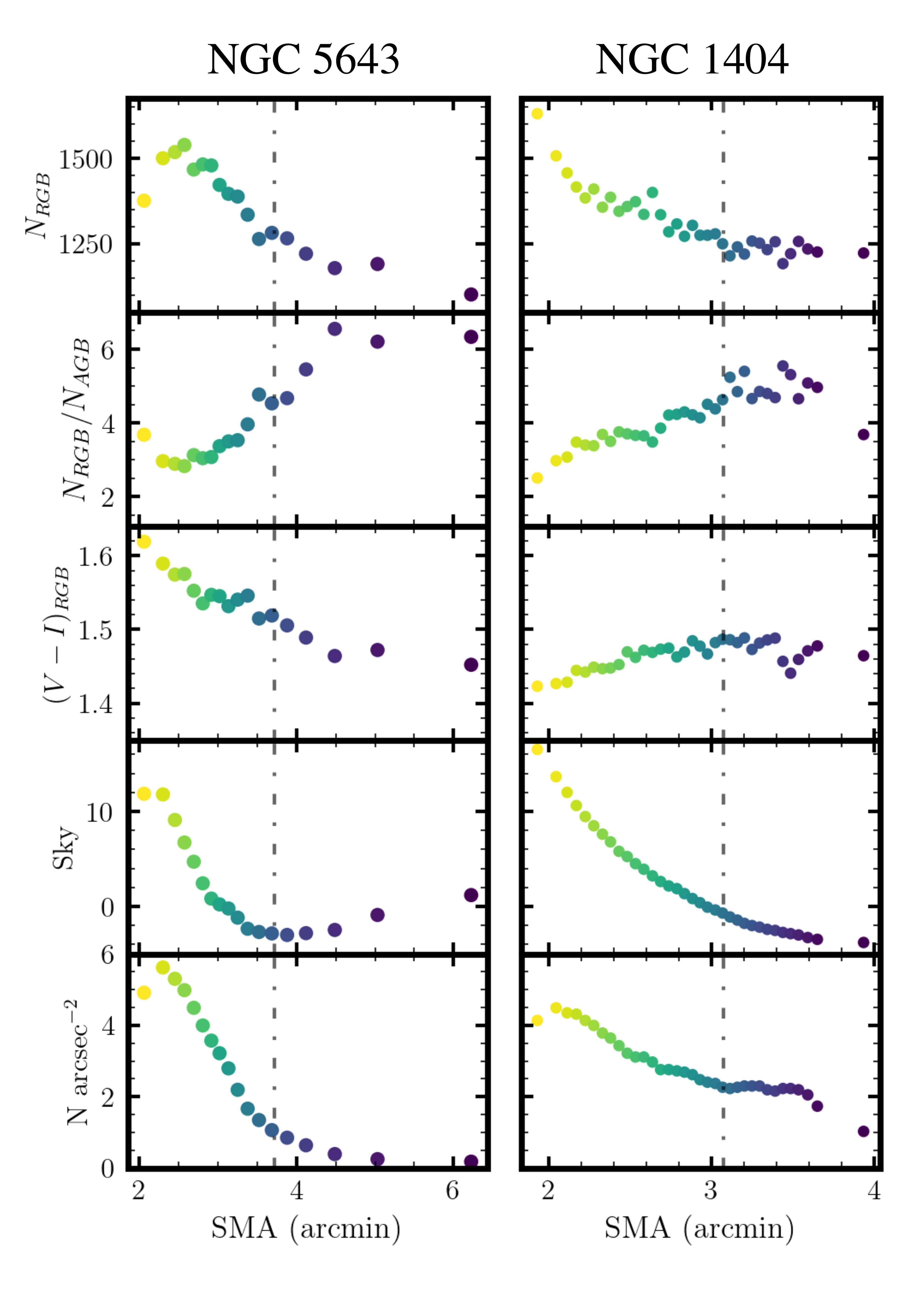}
   \caption{Photometric quantities as proxies for physical parameters for NGC 5643 (left) and NGC 1404 (right) plotted as a function of the mean semi-major axis of each spatial bin. From top to bottom: (i) number of stars classified as RGB, (ii) ratio of RGB to AGB stars, (iii) mean (F606W$-$F814W) color of the RGB stars, (iv) the mean local background, zeroed to the modal sky value for each chip, (v) the average areal source density. The color of each point matches with its corresponding elliptical region, for which a subset are plotted in \autoref{fig:spatial_regions}. The vertical, dot-dashed lines demarcate our adopted ``stellar halo'' boundaries, outside of which we use all sources to determine the TRGB distances.\label{fig:spatial_diag}}
\end{figure*}

For each region, we compute the following statistics: $N_{RGB}$, $N_{AGB}$, $(V-I)_{RGB}$, $Sky_{avg}$, and N arcsec$^{-2}$. Again, $N_{RGB}$ is defined as the number of stars measured to have colors (F606W$-$F814W) $ > 1.00 $ mag and which lie within 1 mag fainter than the adopted TRGB magnitude. Conversely, $N_{AGB}$ has the same color constraint with an oppositely signed magnitude constraint. $(V-I)_{RGB}$ is defined as the average color of the same sources that contribute to $N_{RGB}$ (where $V$ and $I$ are used synonymously with F606W and F814W). The $<Sky>$ is the average local sky counts (in ADU) returned by DAOPHOT for all stars contained in each region, where the values have been zeroed to the modal sky value in each chip. In \autoref{fig:spatial_diag}, we plot all of these quantities as a function of the average semi-major axis of each bin.

\begin{figure}
    \centering
    \includegraphics[width=0.9\columnwidth]{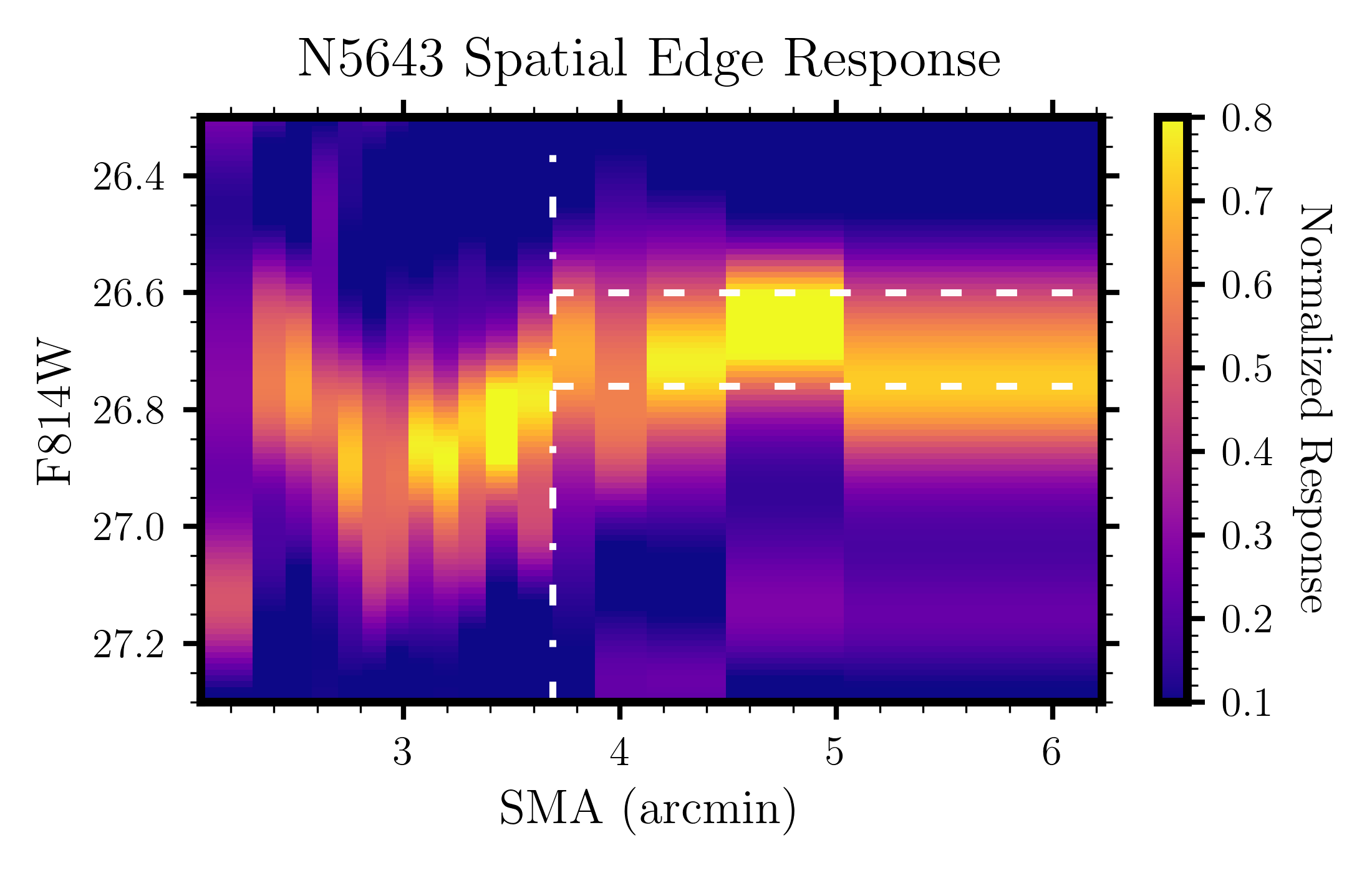}
    \includegraphics[width=0.9\columnwidth]{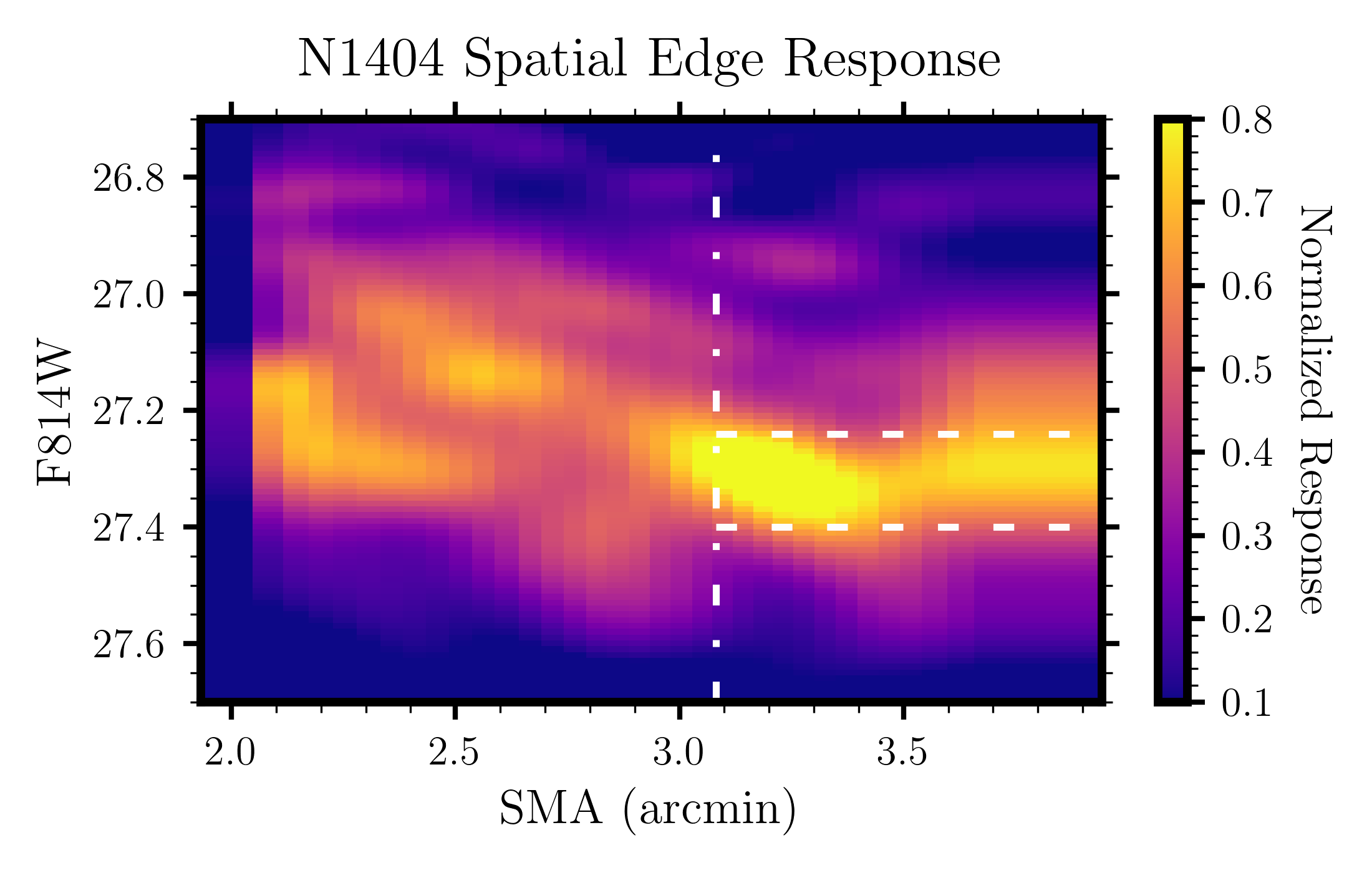}
    \caption{Spatial edge response figures for NGC 5643 (top) and NGC 1404 (bottom). Each vertical bin in semi-major axis represents the first-derivative response for one spatial region, a subset of which are shown in \autoref{fig:spatial_regions}. Each vertical white dot-dashed line represents the ``halo'' boundary adopted (see \autoref{sect:spatial} for an explanation). The white dashed lines represent the 2-sigma confidence intervals on the TRGB magnitudes as determined from the binned Halo CMDs shown in \autoref{fig:detections}. The power in the edge response curve within each bin is normalized, then linearly re-scaled so that the peak value across all bins is equal to one. There are 18 and 35 vertical bins in SMA for NGC~5643 and NGC~1404, respectively. Each SMA slice contains 1000-2000 total stars, with the quantity $N_{RGB} - N_{AGB}$ held constant across all bins.}
    \label{fig:spatial_detect}
\end{figure}

Then, in \autoref{fig:spatial_detect} we simultaneously plot the edge response (the right-most curves plotted in each tri-panel of \autoref{fig:detections}) for all spatial bins at once. The $x$-axis is the newly added spatial dimension (the mean SMA of each region), the $y$-axis is F814W magnitude, and the color map represents the power in the edge response function (corresponding to the $x$-axis in the right-most panels of \autoref{fig:detections}). The response function within each spatial bin is first normalized, then re-scaled such that the maximum value across all bins is unity.

Finally, in the right column of \autoref{fig:detections} we present our finalized TRGB detections for both galaxies. The Halo CMDs contain no young, blue stars, the luminosity functions are well-populated, and the edge responses become unambiguous and single-peaked. Now, we separately describe for each galaxy the details and results of the spatial analysis.

\subsubsection*{NGC 5643}

For NGC 5643, to define the elliptical profile used in the boundary optimization algorithm, we adopted the outermost isophote of the disk component in DSS Blue band imaging (which comprises part of the RGB image shown in \autoref{fig:pointings}) that converged in the \textit{isophote} code in the \textit{astropy} package. The final profile adopted for the spatial analysis is described by position angle $ \phi = 10.14~\pm~1.12 $ deg, ellipticity $e = 0.173~\pm~0.006$, origin (RA, DEC) = (14:32:40.12$~\pm~0.06''$, $-$44:10.0:26.3$~\pm~0.1''$), at  semi-major axis $ a = 0.0608 $ deg. We then run the region boundary optimization described above and generate 18 individual regions, ranging from $ 2' \leq SMA \leq 7' $. In \autoref{fig:spatial_regions}, we show a representative set of these regions plotted over a source density map.

In the left column of \autoref{fig:spatial_diag}, we plot the previously defined parameter values as a function of the mean SMA of each spatial bin. The salient feature common to all panels is a steep gradient for $SMA \lesssim 4'$ and the transition to shallower gradients for $SMA \gtrsim 4'$. The observed stellar population becomes increasingly RGB-dominated, bluer, and the average sky value decreases. We take this to be a transition from disk- to halo- dominated stellar populations.

There is an additional reversal in the Sky values that is not seen in the other parameters. As discussed in \autoref{sect:data}, there was a faint blue glow detected in the NW corner of the HST imaging, which manifests again here. Because this upturn in Local Sky background does not coincide with a decrease in the RGB/AGB fraction, we can conclude that it is not associated with an extragalactic structure, and is mostly likely a result of scattering within the optics from the many bright foreground stars present in the field of view, and potentially a much brighter one that is just off frame. We note that this slight change in the local sky background will not significantly affect the accuracy of our PSF photometry; the increase in sky background simply reduces the depth of the photometry.

In \autoref{fig:spatial_detect}, the power of the response function appears to track tightly with the astrophysical proxy quantities discussed in the previous paragraph. In the inner portions of the disk ($SMA \lesssim 3'$), the TRGB is undetectable. For $ 3' < SMA <  4' $, the peaks in the response functions are significant and faint (F814W $\simeq 26.8$ mag), suggesting the existence of a real TRGB population, but with a higher metallicity or dust content (and accompanying gradients) in those inner regions.
Finally, the peak in the response function begins to stabilize at $SMA > 4' $. Based on the behaviors seen in \autoref{fig:spatial_diag} and \autoref{fig:spatial_detect}, we adopt as our selection of the stellar halo of NGC~5643 all regions that lie outside a semi-major axis equal to 3.9' (14 kpc).

The resulting halo TRGB detection is shown in the top-right panel of \autoref{fig:detections}. From directly comparing the two CMDs, we confirm that our adopted spatial selection has removed both the blue main sequence and red supergiant populations, leaving only what appears to be a predominantly old stellar population. Comparing the edge response functions, the peak in the Halo-based response is brighter, and the long tail to fainter magnitudes, as seen in the full sample's edge response, is significantly reduced in strength. As such, we interpret this tail to fainter magnitudes to be sourced by a reddening gradient associated with the outer edge of the NGC~5643 disk, which is consistent with the gradient in peak location seen in \autoref{fig:spatial_detect} for $3' \leq SMA \leq 3.8'$. Both comparisons tell us that our Halo selection succeeds in avoiding contamination from young stellar populations and minimizes the effects of dust extinction on our TRGB measurement.

We determine the apparent, (foreground) reddened TRGB magnitude for NGC~5643 to be \mtsubscript~=  \spliterr{\mtrgbGALONE}{\measSTATerrGALONEround (stat)}{\measSYSerrGALONEround (sys)}. In \autoref{subsect:artstar} we describe our procedure to estimate the uncertainties in our measurement.

\subsubsection*{NGC~1404}
For NGC~1404, because it is an elliptical galaxy we simply adopt a profile measured directly from the nucleus captured in our ACS imaging, as opposed to fitting for the outermost converged isophote in ground-based imaging, which is what was done for NGC~5643. The final elliptical profile adopted is: position angle $ \phi = 63.0 $ deg, ellipticity $e = 0.063$, and origin (RA, DEC) = (03:38:51.9, -35:35:38.0). As discussed in the previous section, we proceed with only Chip 1 for any analysis, since Chip 2 contained the saturated nucleus of the galaxy and was dominated by unresolved light and blended sources.

Given that crowding effects are still present in the inner regions of this chip, there are a number of caveats to consider when examining \autoref{fig:spatial_diag}.
For example, $N_{AGB}$ and $N_{RGB}$ are defined as the number of stars in intervals that extend one magnitude brighter and fainter, respectively, than the adopted TRGB magnitude. In an uncrowded field, these definitions provide a good estimate of the relative fractions of old ($> 5-10$ Gyr) vs. intermediate-aged ($\sim 1-5 $ Gyr) giant populations. However, in a dense field, crowded RGB stars can be scattered significantly brighter than their uncrowded counterparts, biasing both estimators $N_{RGB}$ and $N_{AGB}$ higher than their true astrophysical values. Fractionally, however, $N_{AGB}$ is biased significantly more than $N_{RGB}$ because $N_{AGB} < N_{RGB}$, and as a result the ratio $N_{RGB}/N_{AGB}$ is biased to smaller values by crowding effects. The estimated color of RGB stars $(V-I)_{RGB}$ will be similarly skewed by blends and biased toward bluer colors, since bright, evolved RGB stars are on average redder than their faint siblings lower on the RGB. This effect can be seen in action in the N1404-C2 panel of \autoref{fig:cmds}, where crowded RGB stars form an unphysical vertical plume that extends to two magnitudes brighter than the TRGB (which can be identified in the N1404-C1 CMD at 27.3~mag) and at bluer colors.

Fortunately, whether the quantities considered in \autoref{fig:spatial_diag} are more impacted by crowding or real astrophysics is not relevant to the conclusions here. After all, the goal is to minimize all systematic errors, either astrophysical (e.g., age, metallicity) or photometric (e.g., crowding). 
Indeed, the photometry parameters suggest that NGC~1404 is more well-mixed than NGC~5643, with shallower gradients in the RGB to AGB ratio and the surface density than those observed in NGC~5643, which is within expectations considering their respective galaxy types.

In \autoref{fig:spatial_detect}, the edge response for $SMA < 3'$ is ambiguous with a forest of edge detection peaks spanning from 27.3 to 26.8 mag, with a general trend towards fainter peak values with increasing $SMA$.
We suspect that these multiple, brighter edges are powered by different levels of crowding affecting the RGB stars located in these inner regions.
We find that $SMA = 3.1'$ is the ideal boundary to adopt for our ``halo'' selection, because $(V-I)_{RGB}$ and $N_{RGB}/N_{AGB}$ attain their peak values and remain mostly stable outside of this boundary.
Most importantly for our purposes, the 2D response function in \autoref{fig:spatial_detect} reaches an unambiguous maximum at this $SMA$, and does not exhibit significant functionality for $SMA > 3.1'$.

In the bottom-left panel of \autoref{fig:detections} we show the TRGB detection from the full sample, and in the bottom-right panel the same for the ``Halo'' selection, both for only the N1404-C1 photometry. Going from the full sample measurement to the Halo measurement, the edge response has significantly sharpened, with the tail out to bright magnitudes significantly reduced. We interpret this to be due to a reduction in the number of crowded RGB stars above the true TRGB magnitude. That would also be consistent with our earlier-discussed interpretation of \autoref{fig:spatial_diag}, where the inner regions of the Chip 1 imaging were seen to exhibit a high enough frequency of source blends so as to bias our estimates of $(V-I)_{RGB}$ and $N_{RGB}/N_{AGB}$.

With a spatial selection locked in, we determine an apparent, (foreground) reddened TRGB magnitude for NGC 1404, \mtsubscript~=\spliterr{\mtrgbGALTWO}{\measSTATerrGALTWOround (stat)}{\measSYSerrGALTWOround (sys)}. In the following section we describe our estimation of the uncertainties in this measurement.

\subsection{Error Estimation with Artificial Stars} \label{subsect:artstar}

\begin{figure*}
    \centering
    \includegraphics[width=0.4\textwidth]{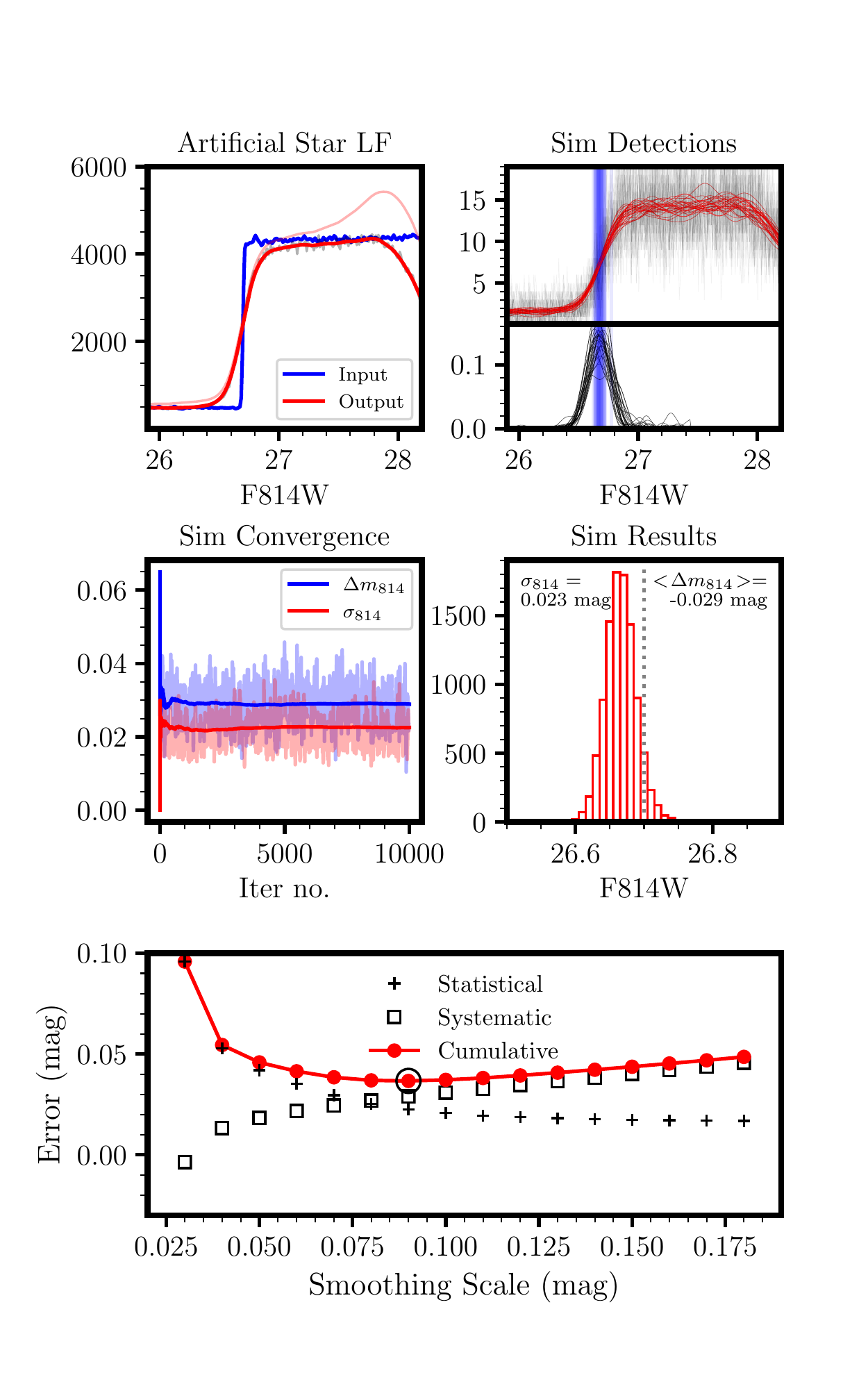}
    \includegraphics[width=0.4\textwidth]{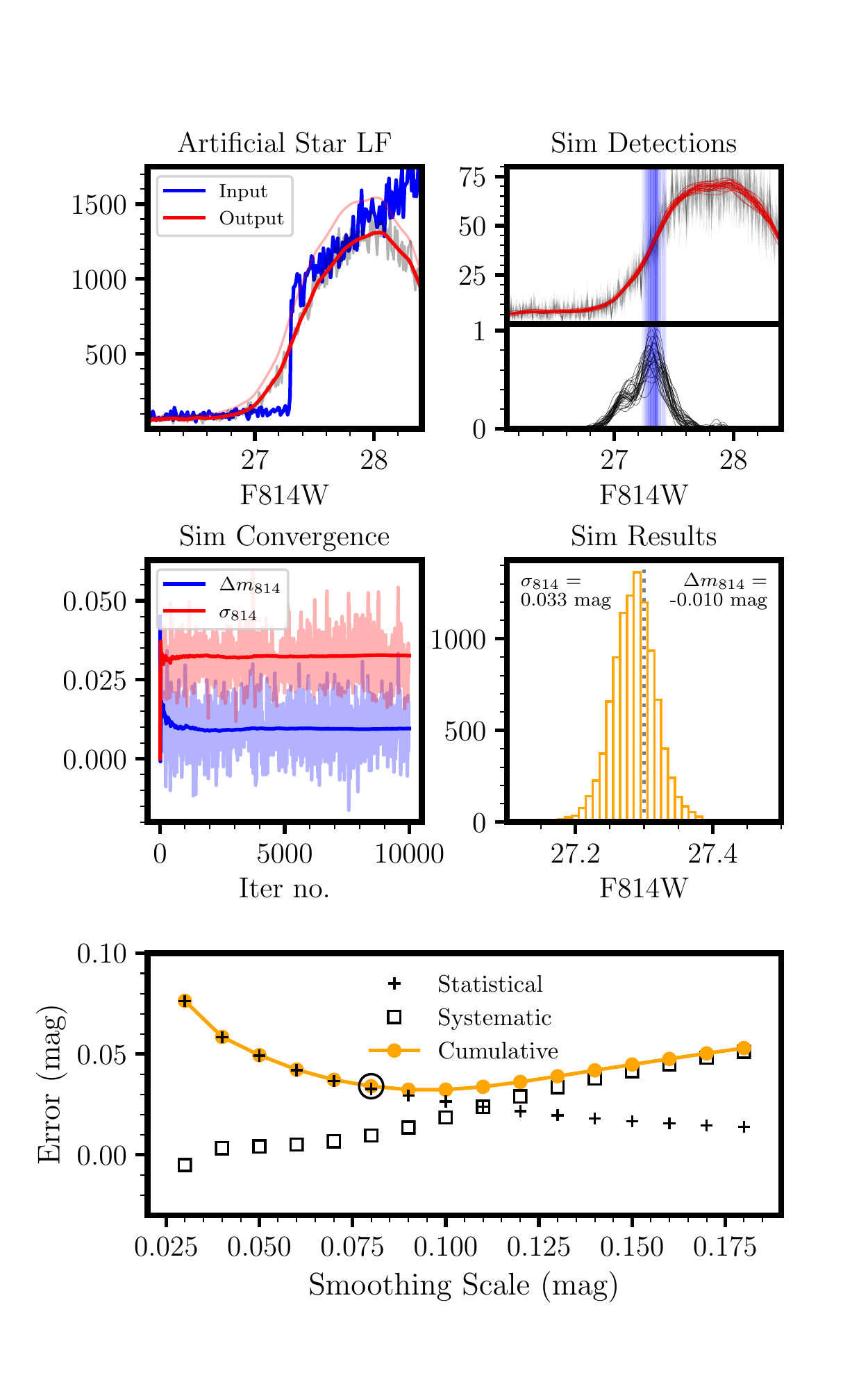}
    \caption{Results of the artificial star edge detection simulations for NGC~5643 (left figure set) and NGC~1404 (right figure set). \textit{Upper-left}: Plotted are both the input and output artificial star luminosity functions (ASLFs), in blue and red, respectively. The vertical steps in each blue curve mark the input TRGB magnitudes 26.7 and 27.3 mag, for NGC~5643 and NGC~1404, respectively. The transparent red curve represents the raw measured LF, while the opaque red curve represents the measured LF after making photometry quality cuts.
    \textit{Upper-right}: Overplot of 30 LFs and simulated TRGB detections, downsampled from the output ASLF to reflect the stellar counts observed in the real data. The edge responses are plotted below.
    \textit{Middle-left}: Plot showing the convergence of our edge detection simulations. The darker curves represent the cumulative average of each quantity, while the transparent curves show the point-to-point variations smoothed by a boxcar of size 25.
    \textit{Middle-right}: Histogram of individual TRGB detections from 10000 downsampled realizations of the full ASLF.
    \textit{Bottom}: Trend of $\sigma_{814}$ and $<\!\Delta m_{{814}}\!>$ as a function of smoothing kernel size. The circled point corresponds to the smoothing scale shown in the Detections, Convergence, and Results plots, and which is adopted to make the TRGB measurement.
    }
    \label{fig:artstar_results}
\end{figure*}

Because our edge detection determination of the TRGB is non-parametric, we turn to simulations of the photometry and edge detection procedure in order to estimate the uncertainties in our TRGB measurements. This procedure was introduced in \citet{hatt_2017} and we summarize it here. We build up a library of $\sim 1$ million artificial stars, which are drawn from a luminosity function (LF) with AGB slope 0.1 dex/mag, and RGB slope 0.3 dex/mag. At one time, we randomly sample 4000 ``artificial stars'' from the model LF and inject them self-consistently into all exposures in a dataset. We then proceed to photometer these injected-source images following the same photometric routines described previously. The injection procedure is then repeated 120 times for each chip, until photometry of at least half a million artificial sources have been tabulated. The input and output (before and after applying photometry quality cuts) artificial star luminosity functions (ASLFs) are shown in the upper-left panels in \autoref{fig:artstar_results}.

The photometry quality cuts (discussed in Section 2.3) are then applied, and the output ASLF is downsampled to reflect as closely as possible the number of stars on either side of the TRGB discontinuity observed in the real data. Thirty of these downsampled ASLFs, and the resulting edge detector response functions are shown in the upper-right panel of \autoref{fig:artstar_results}. This downsampled edge detection procedure is then repeated for 10000 realizations, to ensure convergence in the simulations, defined to occur when the values of the scatter $\sigma_{814}$ and the offset $\Delta m_{{814}}$ reach stable values, as shown in the middle-left panels of \autoref{fig:artstar_results}. 

The final values of these parameters are then adopted as the statistical and systematic errors on our edge detection measurement. The distribution of all 10000 simulated TRGB measurements, as well as the final adopted uncertainties, are shown in the middle-right panel of \autoref{fig:artstar_results}. Additionally, because this process injects sources directly into the images, this evaluation of the uncertainties explores all of the following: crowding biases, biases due to incompleteness, and convolution of the photometric error distribution with the luminosity function.

In the bottom panel of \autoref{fig:artstar_results}, we plot $\sigma_{814}$ and $\Delta m_{{814}}$ as a function of the smoothing kernel width. Circled is the smoothing size adopted for the final TRGB detection. The smoothing size for NGC~5643 was chosen to minimize the cumulative error curve. The smoothing scale adopted for NGC~1404 was chosen to avoid merging the dominant peak observed with adjacent secondary peaks.

In addition to the edge detection uncertainties determined from the artificial star experiments, we also add to the systematic error budget three terms associated with our photometric calibration: (i) the dispersion observed in the frame-to-frame PSF-to-aperture corrections (see \autoref{tab:apcors} for values), (ii) a \EEerr~mag term for the uncertainty in the ACS EE corrections to an infinite aperture, and (iii) a \ZPerr~mag term for the uncertainty in the ACS photometric zero-point. The quadature sum of these three calibration errors and the edge detection systematic error combine to give the $\sigma_{sys}$ column in \autoref{tab:dists}.

\subsection{TRGB Distances} \label{subsect:dists}

\begin{deluxetable*}{ccccccccccccc} 
\tabletypesize{\normalsize} 
\tablewidth{0pt} 
\tablecaption{TRGB magnitudes, foreground reddenings, and true distances to NGC 5643 and NGC 1404. \label{tab:dists}} 
\tablehead{ 
\colhead{Galaxy} &
\colhead{\mtsubscript} & 
\colhead{$\sigma_{stat}$} &
\colhead{$\sigma_{sys}$\tablenotemark{a}} &
\colhead{$A_{814}$} &
\colhead{$\sigma_{A_{814}}$} &
\colhead{$\left(m-M\right)_0$\tablenotemark{b}} &
\colhead{$\sigma_{stat}$} &
\colhead{$\sigma_{sys}$\tablenotemark{a}} &
\colhead{$D$} &
\colhead{$\sigma_{stat}$} &
\colhead{$\sigma_{sys}$} & \\
\colhead{} &
\colhead{(mag)} &
\colhead{(mag)} &
\colhead{(mag)} &
\colhead{(mag)} &
\colhead{(mag)} &
\colhead{(mag)} &
\colhead{(mag)} &
\colhead{(mag)} &
\colhead{(Mpc)} &
\colhead{(Mpc)} &
\colhead{(Mpc)} &
}
\startdata 
NGC 5643 & \mtrgbGALONEround & \measSTATerrGALONEround & \measSYSerrGALONEround & \ACSIextinctionGALONEround & \ACSIextinctionerrGALONEround & \distmodGALONEround & \totSTATerrGALONEround & \totSYSerrGALONEround & \distGALONEround & \distSTATerrGALONEround & \distSYSerrGALONEround \\
NGC 1404 & \mtrgbGALTWOround & \measSTATerrGALTWOround & \measSYSerrGALTWOround & \ACSIextinctionGALTWOround & \ACSIextinctionerrGALTWOround & \distmodGALTWOround & \totSTATerrGALTWOround & \totSYSerrGALTWOround & \distGALTWOround & \distSTATerrGALTWOround & \distSYSerrGALTWOround \\
\enddata
\tablenotetext{a}{Includes uncertainties in PSF-to-aperture corrections, encircled energy corrections to an infinite aperture, and the uncertainty in the STScI infinite aperture zero-point.}
\tablenotetext{b}{$M_{\text{814}}^{\mathrm{TRGB}}=$\spliterr{\trgblum} {\trgblumSTATerr ~(stat)} {\trgblumSYSerr ~(sys)}~mag \citep{freedman_2020}.}
\end{deluxetable*} 

In this section, we first consider the effects of a color cut on our observed TRGB magnitudes before determining the true TRGB distances to NGC~5643 and NGC~1404 via correction for foreground reddening and application of the CCHP calibration of the TRGB luminosity.

We tested the effects of a color-selection on our measured TRGB magnitudes. This selection, also considered in previous CCHP papers, is defined by slopes $-4.00$~mag/mag, and bounded by $ 0.90 < $ (F606W$-$F814W) $ < 2.00 $ mag at the magnitude of the TRGB. Additionally, we perturbed the data quality cuts described in \autoref{sect:data} and found the effects of both on the detected TRGB magnitude to be negligible ($<0.01$ mag).

The apparent TRGB magnitudes $m_{814}^{\mathrm{TRGB}}$ are corrected for foreground reddening using the \citet{schlafly_2011} recalibration of the \citet{schlegel_1998} dust maps and their preferred normalization of a \citet{fitzpatrick_1999} reddening law with $R_V=3.1$. Next, we apply the CCHP calibration of the TRGB absolute magnitude $M_{814} =$ \spliterr{\trgblum}{\trgblumSTATerr (stat)}{\trgblumSYSerr (sys)} \citep{freedman_2020} to the de-reddened TRGB magnitude to determine the true distance to each galaxy.

For NGC~5643, this results in an extinction in the F814W band $A_{814} = \ACSIextinctionGALONE \pm \ACSIextinctionerrGALONE $ mag, corresponding to $E(B-V) = 0.15 \pm 0.02 $ mag, where we have conservatively rounded up to the nearest hundredth of a magnitude the 10\% (0.015~mag) uncertainty suggested by \citet{schlafly_2011} for observations made near the galactic plane. Applying the CCHP calibration, we find for NGC 5643 a true distance modulus, $\mu_0 = $\spliterr{\distmodGALONEround}{\totSTATerrGALONEround (stat)} {\totSYSerrGALONEround (sys)}

For NGC~1404, we again correct for the foreground reddening measured by maps presented in \citet{schlegel_1998} and recalibrated by \citet{schlafly_2011}. However, since NGC~1404 does not lie near the galactic plane, we cannot adopt the 10\% uncertainty suggested by \citet{schlafly_2011}. Instead, we conservatively adopt half the measured reddening as the error on the measurement, i.e., $E(B-V) = 0.01 \pm 0.005$ mag, or $A_{814} = \ACSIextinctionGALTWO \pm \ACSIextinctionerrGALTWO $ mag. Finally, with the \citet{freedman_2020} zero-point, we find for NGC~1404 a true distance modulus, $\mu_0 = $\spliterr{\distmodGALTWOround}{\totSTATerrGALTWOround (stat)} {\totSYSerrGALTWOround (sys)}. 

The above measurements and uncertainties are summarized in \autoref{tab:dists}.

\section{Impact on the CCHP SN~Ia Calibration} \label{sect:snia_calib}

To measure the Hubble Constant, it is necessary to measure distances and velocities to objects that extend sufficiently far into the low-redshift ($z \leq 0.2 $) Hubble Flow, where peculiar velocities are negligible compared to cosmological redshifts, but where $\Delta H(z)$ remains vanishingly small. The current gold standard to accomplish this is through the use of SNe~Ia, which demonstrate a remarkably small dispersion ($\sigma \le 0.10$ mag) in the Hubble Diagram, after correcting for the shapes of their light curves. However, because SNe~Ia are relatively rare (one per galaxy per century), none have recently occurred nearby enough to measure a direct physical distance. The hope is that this can change with accurate measurement of time delays to multiply lensed SNe~Ia \citep[e.g.,][]{refsdal_1964}. But as of now, the only way to determine the intrinsic luminosity of the SNe~Ia is to measure distances to SN~Ia host galaxies using a primary distance indicator (e.g., Cepheids or the TRGB).

\begin{figure*}
    \centering
    \includegraphics[width=0.75\textwidth]{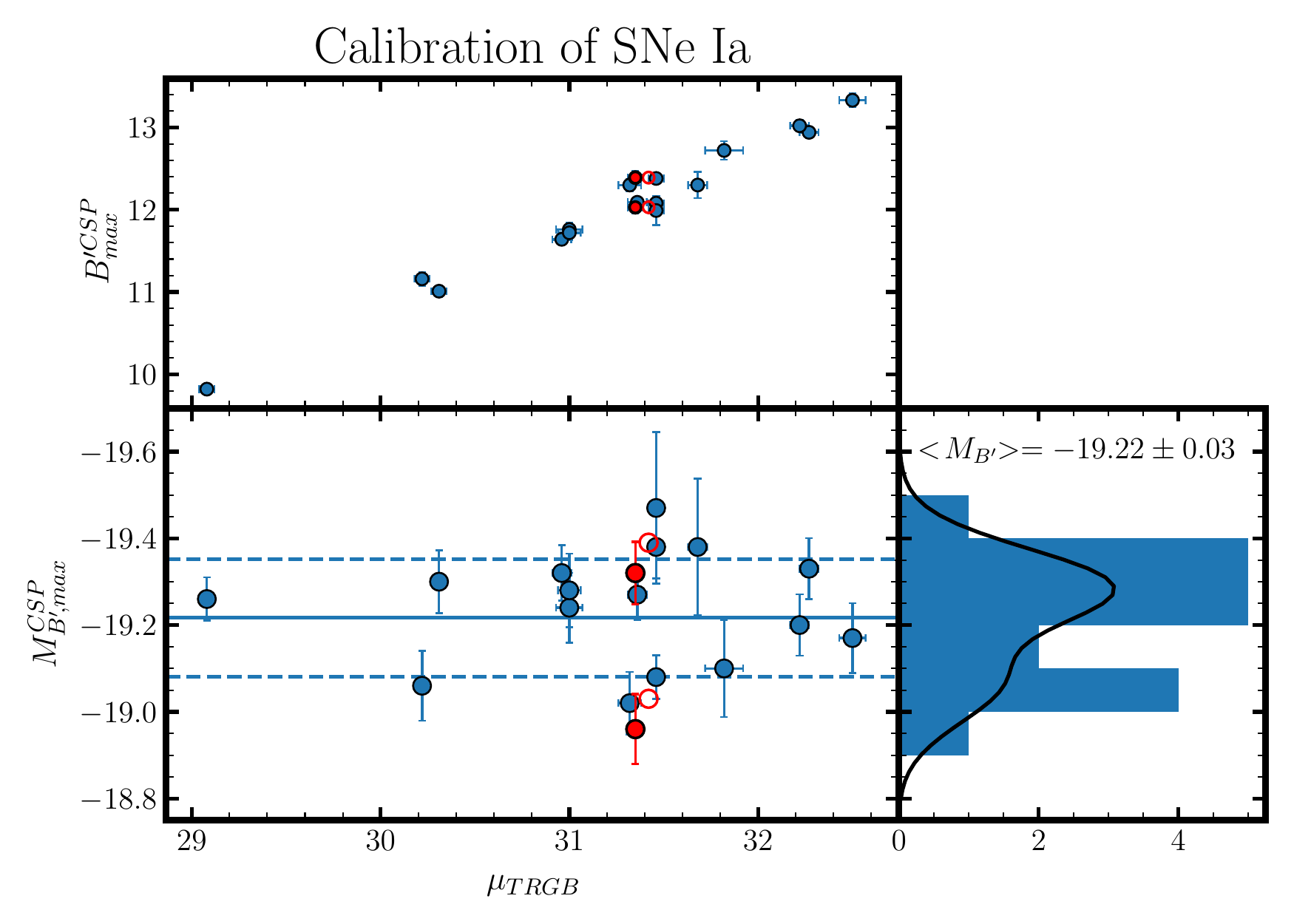}
    \caption{Updated CCHP calibration of the CSP Type Ia supernovae distance scale. Blue circles represent all distances and SN magnitudes presented in Table 3 of \citet{freedman_2019}, except for NGC~1404. Red circles are used 
    to distinguish NGC~1404 from the rest of the sample, where the original \citeauthor{freedman_2019} values are plotted as open circles, while the new values (based on the TRGB distance presented in this study) are plotted as red, filled circles. The dashed lines represent the 1-$\sigma$ standard deviation of the sample.
    }
    \label{fig:snia_calib}
\end{figure*}

In the first CCHP results paper \citep{freedman_2019}, ten TRGB distances to SN~Ia host galaxies were combined with five others determined by \citet{Jang_2017a}. In total, 15 TRGB distances and 18 SNe Ia were used to absolutely calibrate the SN Ia distance scale. One of the TRGB distances included was that to NGC~1404, which was taken to be the average of the distances to NGC~1316 and NGC~1365. And here we want to update the CCHP calibration of the SN~Ia luminosity with the newly measured TRGB distances presented in this study. 

Normally, in order to incorporate new distances and SNe into the calibration of the SN~Ia distance scale, one would need to re-fit the full cosmological SN~Ia dataset, simultaneously re-determining the best-fit parameters to correct for the shapes of the SN light curves. However, because NGC~1404 was already included in the \citeauthor{freedman_2019} analysis, we can bypass this requirement, update its TRGB distance, and immediately compute a new SN~Ia absolute magnitude. By contrast, we cannot do the same for NGC~5643 since it was not included in our first $H_0$ results paper, and refitting the entire SN~Ia dataset is beyond the scope of this paper.

Shown in the top panel of \autoref{fig:snia_calib} is the corrected B-band peak magnitude (the standardize-able candle of the SN~Ia distance scale) $B'^{CSP}_{max}$ vs. TRGB distance modulus $\mu_{TRGB}$.\footnote{$B'^{CSP}_{max}$ is defined in Equation 2 of \citet{freedman_2019}.} 
Blue points and open red circles are values taken directly from Table 3 of \citet{freedman_2019}, where the open red circles represent the values based on the old, averaged distance to NGC~1404. The filled red circles, on the other hand, represent the updated direct distance to NGC~1404 presented here. Plotted in the bottom panel is the resultant absolute magnitude $M_{B',max}^{CSP} = B'^{CSP}_{max} - \mu_{TRGB}$ as a function of $\mu_{TRGB}$, where the points follow the same scheme as in the top panel.

In the right panel we show the $M_{B',max}^{CSP}$ values marginalized over the $\mu_{TRGB}$ axis. This histogram presents the calibration of the SN~Ia luminosity, which sets the physical scale for the SNe~Ia measured out in the Hubble Flow.
The value of the updated SN~Ia calibration is printed immediately above the histogram, and is within 0.01 mag of the value found in \citet{freedman_2019}. We have included both SNe hosted by NGC~1404 in our re-calculation of the SN Ia absolute magnitude, but see \autoref{sect:1404_appendix} for a discussion on why excluding the fainter SN~2007on from the SN calibrator sample may be necessary.

\section{Comparison with Previous Distance Measurements \label{sect:discussion}}
\subsection{NGC~5643} \label{subsect:5643_dists}
A TRGB measurement determined in NGC~5643 using the same data presented here was released to the Extragalactic Distance Database (G. Anand, priv. comm.).\footnote{\url{http://edd.ifa.hawaii.edu/}}They found an apparent $m_{\mathrm{TRGB}}^{\mathrm{814}} = 26.70 \pm 0.03 $ mag, in agreement with our determination of the same value $m_{\mathrm{TRGB}}^{\mathrm{814}} =$\spliterr{\mtrgbGALONE}{\measSTATerrGALONEround (stat)}{\measSYSerrGALONEround (sys)} mag. Besides that, eight distances have previously been measured to NGC~5643, according to the compilation on the NASA/IPAC Extragalactic Database (NED; \url{https://ned.ipac.caltech.edu/}). All were determined using the Tully-Fisher relation, with a mean equal to 30.25 mag and standard deviation 0.44 mag. 

More recently, however, a number of distances have been published to SN~2017cbv, the most recent of the two SNe hosted by this galaxy. \citet{sand_2018} determined a distance $\mu = 30.45 \pm 0.09$~mag assuming an $H_0 = 72$\hounits. \citet[B20,][]{burns_2020} tested three different light curve fitters on their CSP photometry of SN~2013aa and SN~2017cbv, finding a range of values between 30.62~mag and 30.39~mag. Note that the two extremal values are from one light curve fitter (SALT2) and that the other four distances exhibited a full range of only 0.10~mag. \citet{wang_2020} considered three different methods of determining the distance to SN~2017cbv (all assuming an $H_0 = 72$\hounits): (1) NIR ``calibrations'', (2) SNooPy \citep{burns_2011, burns_2018} light curve fits, and (3) color-magnitude information in the tail of the Ia light curve. First, from considering many different NIR luminosity-width relations they found distances ranging from 30.11 to 30.41~mag. From the SNooPy fits to their optical and NIR photometry, they found $30.46 \pm 0.08$ mag, which is in perfect agreement with the SNooPy fit made to independent CSP photometry. From their late-time color-magnitude fit, they determined $30.58 \pm 0.05$~mag. 

For our purposes, however, we want to know if the SNe~Ia in NGC~5643 are consistent with the \citet{freedman_2019} results. Since that paper used CSP SNe to measure the Hubble Constant, we will consider for the rest of this section the B20 SNooPy distances to 2013aa and 2017cbv. B20 found $\mu_{2013aa}^{72} = 30.47 \pm 0.08$ mag and $\mu_{2017cbv}^{72} = 30.46 \pm 0.08$ mag where the superscript denotes the adopted $H_0$ value of 72 \hounits in their light curve fitter. However, a direct comparison between our TRGB distance and the B20 distances to NGC~5643 would implicitly fold in the assumption of an $H_0$ value that does not agree with that found in \citet{freedman_2019}. To avoid this unlike comparison, we will re-scale the B20 distances so that they are instead calibrated by the existing CCHP TRGB-based $H_0$. We describe this process in more detail in \autoref{app:snia}.

This results in the CCHP-anchored SN distances $\mu_{2013aa}^{69.6} = 30.54 \pm 0.09$ mag and $\mu_{2017cbv}^{69.6} = 30.53 \pm 0.09$ mag, which can be directly compared with the TRGB distance measured here, $\mu_{\mathrm{TRGB}} = $\spliterr{\distmodGALONEround}{\totSTATerrGALONEround(stat)}{\totSYSerrGALONEround(sys)}~mag. The agreement between the TRGB and SN distances indicates that the TRGB distance to, and SNe hosted by, NGC~5643 are consistent with the results presented in \citet{freedman_2019,freedman_2020}.

\subsection{NGC~1404} \label{subsect:1404_dists}
\begin{figure}
    \centering
    \includegraphics[width=0.9\columnwidth]{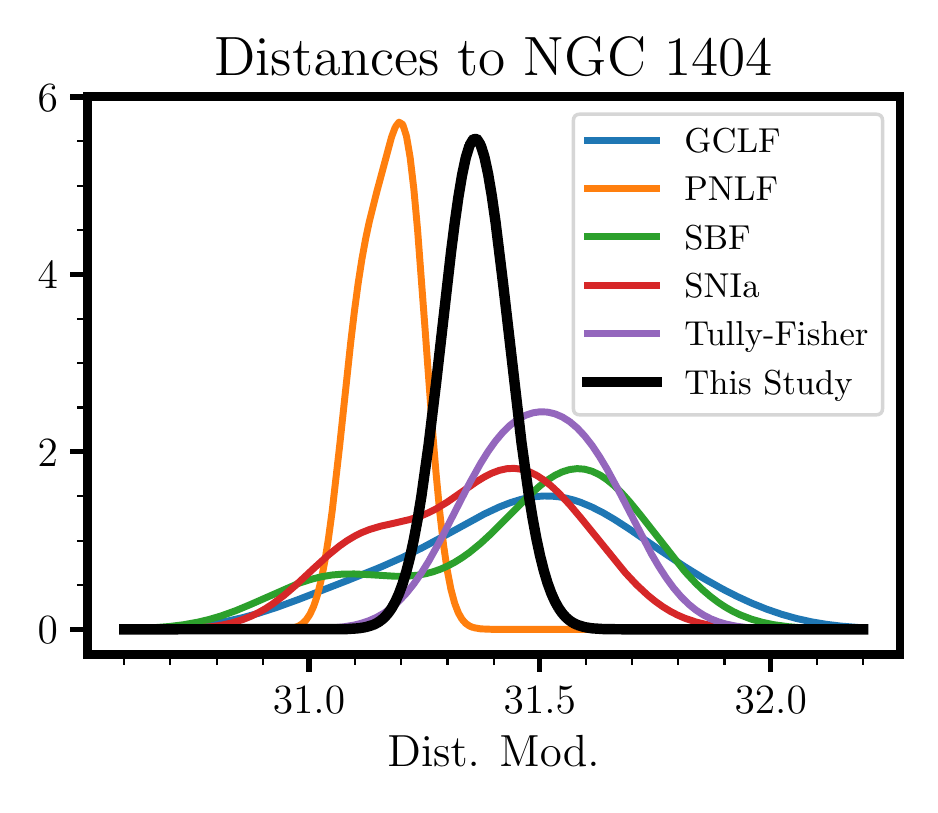}
    \caption{Compilation of previous distances measured to NGC~1404 for the five methods that displayed a standard deviation less than 0.3 mag. The TRGB distance measured in the present study is plotted as solid black. Values acquired from NED.}
    \label{fig:dist_comp}
\end{figure}

For NGC~1404, NED has compiled 55 prior distances coming from ten different distance indicators. In \autoref{fig:dist_comp}, we plot Gaussian-kernel-smoothed distributions for those methods with distances that exhibited standard deviations less than 0.3 mag: GCLF, PNLF, SBF, SN~Ia, and the Tully-Fisher relation (see \autoref{sect:intro} for a full list of references). For comparison, we plot a Gaussian representation of the TRGB distance determined here. There is a large variation in distance measurements, ranging from 31.2 mag (PNLF) to 31.6 mag (SBF), with our TRGB distance landing in the middle. This is consistent with the findings in \citet{hatt_2018b}, who saw an almost identical distribution of distances to NGC~1316. We emphasize that these distances have been determined with methods that do not have a simple geometric calibration like the TRGB or the Cepheids, and, as a result, their absolute zero-points and associated systematic errors are less well-constrained.

Thus, to get a better understanding for where our TRGB measurement stands, we compute the \emph{relative} line-of-sight (for the remainder of this section the term ``distance'' refers to a line-of-sight distance, unless otherwise specified) distances $\Delta \mu = \mu^{1404} - \mu^{1316}$ between NGC~1316 and NGC~1404, as determined with the TRGB \citep[this study and][]{freedman_2019}, SN~Ia \citep{burns_2020}, and SBF \citep[B09,][]{Blakeslee_2009} methods. Both galaxies are massive ellipticals located in the Fornax Cluster, with comparable masses and multiple SNe~Ia to compare. The use of these similar galaxies, should, to some degree, cancel out systematic uncertainties associated with each method of distance measurement. Further reducing the contribution of systematic uncertainties, each set of distances we have chosen to use were determined homogeneously in the same study. 

\begin{figure}
    \centering
    \includegraphics[width=0.9\columnwidth]{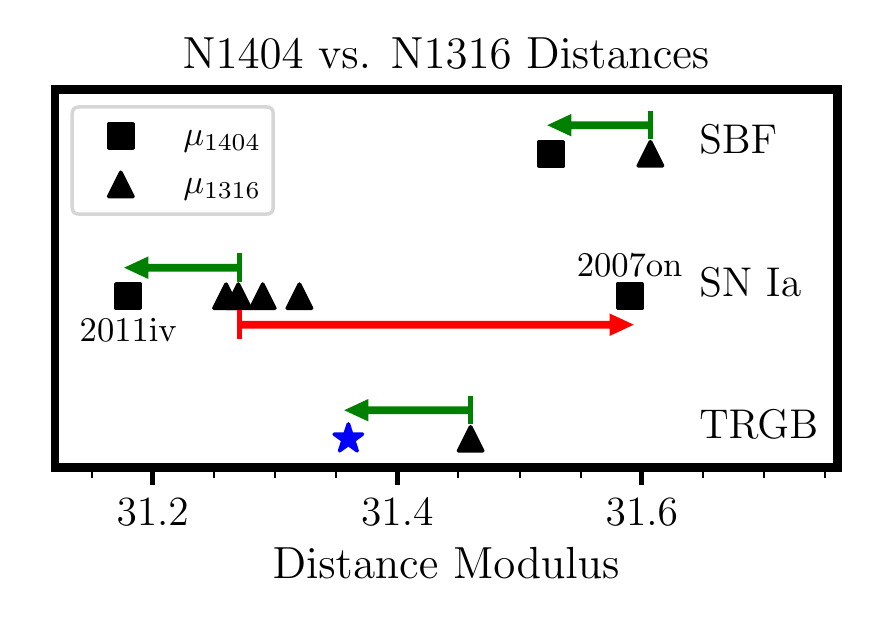}
    \caption{TRGB, SBF, and SN~Ia distances measured to NGC~1404 (black squares) and NGC~1316 (black triangles) in the Fornax Cluster. The vertical axis is arbitrary. We plot as a blue star the TRGB distance to NGC~1404 presented in this study. To compute the pair of relative SN~Ia distances, we use a combined SN distance to NGC~1316 (see text), and consider separately each of SN~2007on and SN~2011iv (individually labeled) in NGC~1404. The direction/color and size of each arrow corresponds to the sign and magnitude of the quantity $\Delta \mu = \mu_{1404} - \mu_{1306}$, where negative values are plotted as green, left-facing arrows, and positive ones as red, right-facing arrows. The CCHP TRGB distance between NGC~1404 and NGC~1316 agrees well with the same as determined using SBFs and SNe (when excluding the subluminous SN~2007on).
    }
    \label{fig:diff_mod}
\end{figure}

The resulting relative distances are: $\Delta \mu_{TRGB} = \FOURTEENdiffTHIRTEENround \pm \FOURTEENdiffTHIRTEENerrround $ mag, $\Delta \mu_{SBF} = -0.081 \pm 0.097 $ mag, $\Delta \mu_{07on} = 0.309 \pm 0.059 $ mag, and $\Delta \mu_{11iv} = -0.101 \pm 0.063 $ mag (see \autoref{sect:1404_appendix} for a detailed description of the calculations). We have computed the line-of-sight SN distances separately for each of 2007on and 2011iv because they are $\sim \! 0.4$~mag discrepant. In \autoref{fig:diff_mod}, we plot these relative distance values as arrows, and the corresponding absolute distances used to calculate them as black squares (NGC~1404) and black triangles (NGC~1316). From comparing the direction and size of the relative distances, it is clear that SN~2007on is significantly underluminous (see \autoref{sect:1404_appendix} for a discussion on this), while SN~2011iv gives a relative distance in very good agreement with the SBF and TRGB relative distances. 

Converting the TRGB-based relative line-of-sight distance $\Delta \mu_{TRGB}$ to physical units, we have $\Delta D_{TRGB} = -0.92 \pm 0.54$~Mpc. If we also include the on-sky separation of the two galaxies, we determine a three-dimensional separation $\Delta r_{TRGB} = \left| \vec{r}_{1404} - \vec{r}_{1316} \right| = 1.50^{+0.25}_{-0.39}$~Mpc. Combining our TRGB-only constraint with the two other consistent relative distance determinations, we compute $\Delta r_{\mathrm{All}} = 1.48^{+0.18}_{-0.25}$~Mpc, providing a powerful constraint on the 3D alignment of these two galaxies within the Fornax Cluster.

Based on the consistency seen in the relative SBF and TRGB distances, our high-precision TRGB distances to these two giant elliptical galaxies should also provide a useful zero-point calibration for the red-edge of the SBF method. We find a combined (NGC~1404 and NGC~1316) offset between our TRGB distances and the B09 (again chosen here for their internal consistency) SBF distances equal to $\mu_{TRGB} - \mu_{SBF} = -0.156 \pm 0.067 $~mag, where we have used the total (stat.+sys.) uncertainties quoted for the TRGB and SBF distances.

\section{Summary and Conclusions} \label{sect:conclusion}
We have presented here point-source photometry from HST ACS/WFC imaging of the galaxies NGC~5643 and NGC~1404. We partitioned the photometry into elliptical subregions to probe how our measurement of a TRGB distance depended on location within each galaxy. From that analysis, we isolated Pop II RGB stars in the outer regions of both galaxies which were then used to determine the distances to NGC~5643 and NGC~1404. For NGC 5643, we determined true modulus \( \mu_0 = \) \spliterr{\distmodGALONEround}{\totSTATerrGALONEround (stat)}{\totSYSerrGALONEround (sys)} and for NGC 1404, \( \mu_0 = \) \spliterr{\distmodGALTWOround}{\totSTATerrGALTWOround (stat)}{\totSYSerrGALTWOround (sys)}. These are the first distances to be measured to these galaxies with a primary distance indicator, and thus the most accurate and precise available.

We also related our new NGC~1404 distance with the existing CCHP distance to NGC~1316 (also a member of the Fornax Cluster), and included in the comparison two homogeneous sets of SN~Ia and SBF distances. We found overwhelming agreement between the \emph{relative} distances as determined with each of the three methods, providing a high-significance constraint on the three-dimensional separation between the two galaxies. At the same time, we found that the absolute SBF distances are systematically fainter/more distant than both the SNe~Ia and TRGB distances to each galaxy, which would appear to confirm an offset between SBF and TRGB distances that was previously found for NGC~1316 alone.

We considered the impact that these new distances would have on our CCHP calibration of the SN~Ia distance scale and, as a result, our measurement of the Hubble Constant. The new TRGB distance to NGC~5643 presented here was found to be in agreement with the Carnegie Supernova Project (CSP) distances measured to its sibling SNe~Ia, and consistent with the conclusions in \citet{freedman_2019}. The distance to NGC~1404 was updated and we re-determined the SN~Ia absolute magnitude to be within 0.01 mag of the value measured by \citet{freedman_2019}. 

\acknowledgments
We thank Peter B. Stetson for his continued provision of, and support for, the DAOPHOT+ALLFRAME photometry software. 
We acknowledge Dr. Dylan Hatt for helpful comments on this manuscript.
TJH thanks Dr. Chris Burns for enlightening discussions on SNe~Ia.
TJH is grateful to the Brinson foundation for their support of his graduate studies.
MGL was supported by the National Research Foundation grant funded by the Korean Government (NRF-2019R1A2C2084019).
Support for program HST-GO-15642 was provided by NASA through a grant from the Space Telescope Science Institute, which is operated by the Association of Universities for Research in Astronomy, Inc., under NASA contract NASA 5-26555.
RLB was supported by NASA through Hubble Fellowship grant \#51386.01 awarded by the Space Telescope Science Institute.
This research has made use of the NASA/IPAC Extragalactic Database (NED), which is operated by the Jet Propulsion Laboratory, California Institute of Technology, under contract with the National Aeronautics and Space Administration.
The imaging data in this paper were obtained from the Mikulski Archive for Space Telescopes (MAST).
We thank the Observatories of the Carnegie Institution for Science and the University of Chicago for their support of our research on the expansion rate of the universe.
\facility{HST(ACS)}
\software{DAOPHOT/ALLFRAME \citep{stetson_1987,stetson_1994}, TinyTim \citep{krist_2011}, astropy \citep{astropy_2013}}

\newpage
\appendix

\section{Objects of Interest in the ACS Imaging}
\subsection*{Globular Cluster Candidates}
\begin{figure}
    \centering
    \includegraphics[width=0.85\columnwidth]{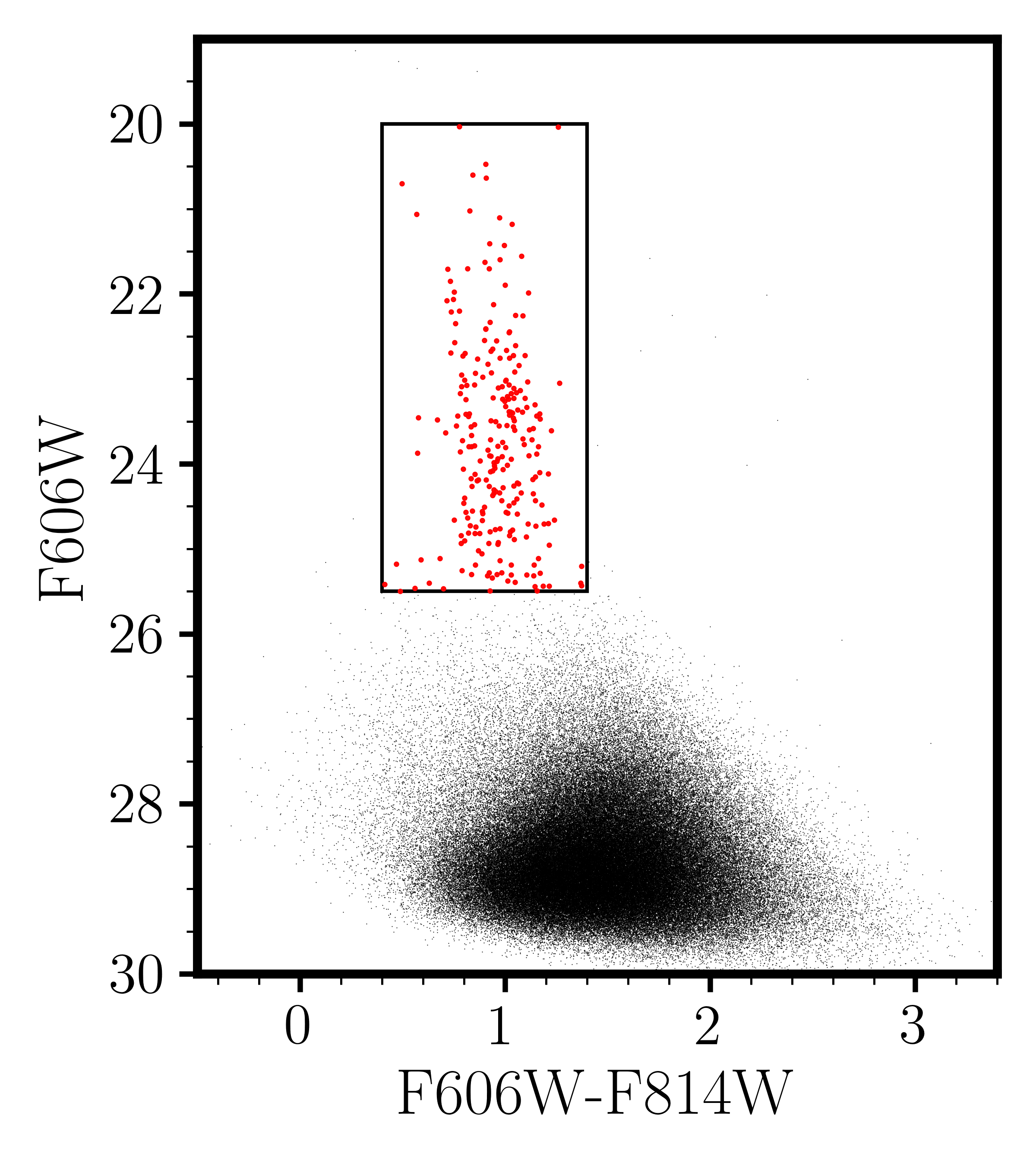}
    \caption{Globular Cluster (GC) candidates in the HST/ACS imaging of NGC~1404. Plotted is the CMD for both Chips 1 and 2 combined. Red points denote the GC candidate sources selected via the plotted selection (black box); they form an expected vertical sequence centered on (F606W$-$F814W) $ = 1.00 $~mag and extend from $V=25$ to $V=20$ mag.}
    \label{fig:gcs}
\end{figure}

Because they share similar RGB-dominated stellar populations, the integrated light from globular clusters (GCs) occupies a narrow strip in photometric color. As a result, GC candidates can be classified with a simple color-magnitude selection. We found that the expected morphology of the GC sequence \citep{forbes_1998} was best observed in a F606W vs. F606W$-$F814W CMD. We combined the N1404-C1 and N1404-C2 CMDs shown in \autoref{fig:cmds} and made the following cuts $ 0.40 < $ (F606W$-$F814W) $ < 1.40 $, and $ 20.0 < $ F606W $ < 25.5 $. The results of the selection are shown in \autoref{fig:gcs}.

Conceivably, the TRGB distance we measured to NGC~1404 could be compared to a distance measured from the globular cluster luminosity function (GCLF) presented here. That comparison could provide a useful evaluation on how the GCLF performs as a distance indicator.

\subsection*{FCC~B1281}
There is a clear resolved dwarf galaxy in the outer region of our HST imaging of NGC~1404. To the best of our knowledge, \citet{ferguson_1989} first documented this object in their Fornax Cluster Catalog (FCC) and gave it the identifier FCC-B1281. It was included there as part of the ``possible members and likely background galaxies'' sub-catalog. Then, \citet{Hilker_1999} definitively classified this object as a likely dwarf member in their Catalog of Galaxies of Fornax (CGF), giving it the identifier CGF~1-44. Most recently, \citet{venhola_2018} observed this dwarf in their Fornax Deep Survey (FDS), naming it FDS11\_DWARF186 within their own catalog of member galaxies.

\begin{figure*}
    \centering
    \includegraphics[width=0.85\textwidth]{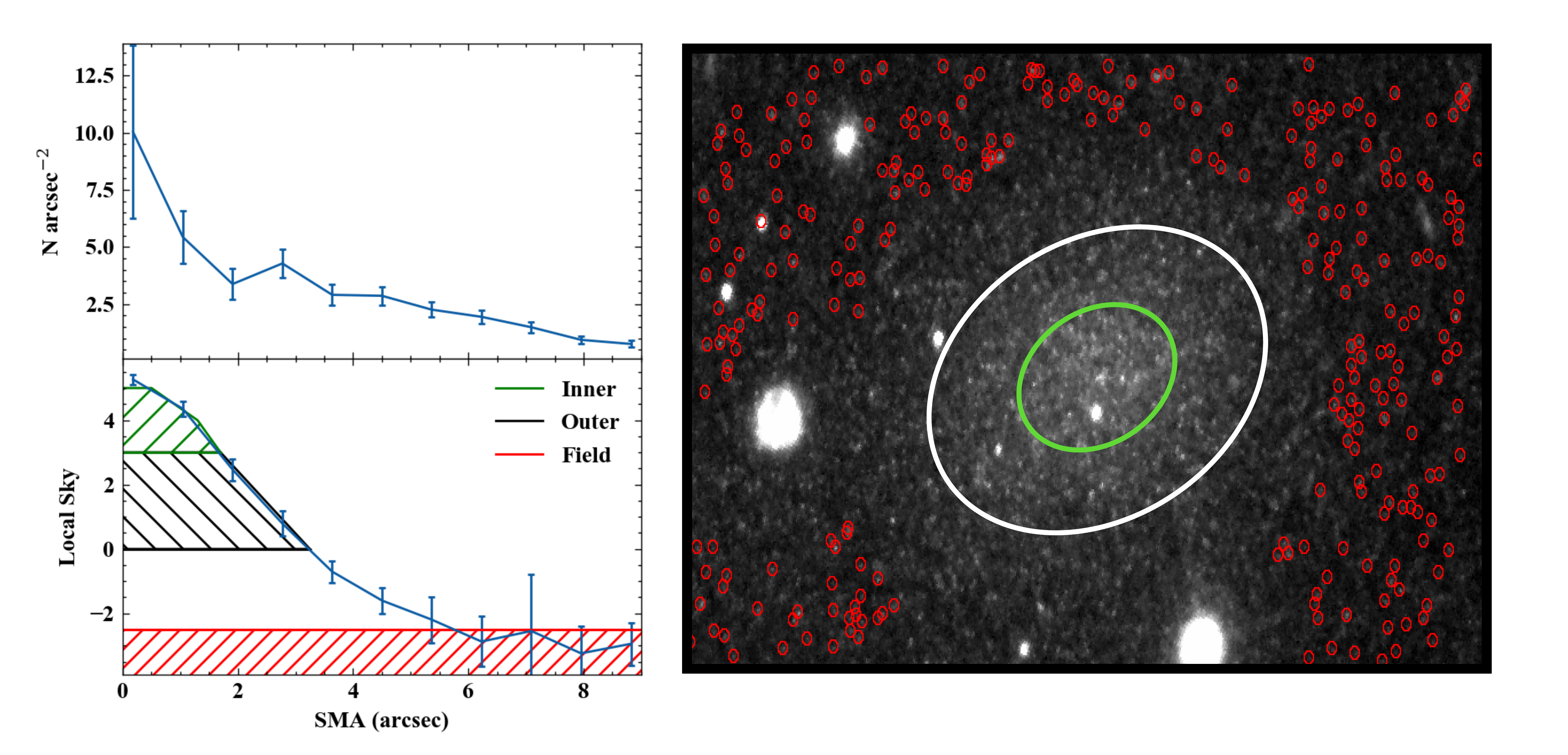}
    \caption{\textbf{Include an RA/DEC grid} Selection of member stars of FCC~B1281. We define three separate regions: Inner (green),  Outer (white/black), and Field (red), where ``Field'' sources are selected to be representative of the main body of NGC~1404.
    }
    \label{fig:dwarf1}
\end{figure*}

It is an interesting question to ask whether we can measure a TRGB distance to this resolved Dwarf galaxy, both to determine whether FCC~B1281 is bound to NGC~1404 and to provide a check on our TRGB distance measured to NGC~1404. To do this, we must first construct a CMD of the Dwarf's stellar population by selecting sources that are sufficiently close-in to the dwarf's center -- to minimize contributions from NGC~1404 -- while also avoiding source blends in its crowded inner regions. To that end, we used the sky value returned by DAOPHOT to construct a surface brightness profile of the galaxy (see bottom-left panel of \autoref{fig:dwarf1}). Within that profile, we defined three regions ``Inner,'' ``Outer,'' and ``Field'' for Local Sky values $ Sky > 3 $, $0 < Sky < 3$, $Sky < -2.5$. As before, the $Sky$ counts here have been zeroed to the modal value across the entire chip. The results of these selections are (de)projected onto the sky in the right-hand panel of \autoref{fig:dwarf1}. There, the Inner and Outer boundaries are marked by green and white ellipses, respectively, while ``Field'' sources are individually identified with red circles.

\begin{figure}
    \centering
    \includegraphics[width=0.9\columnwidth]{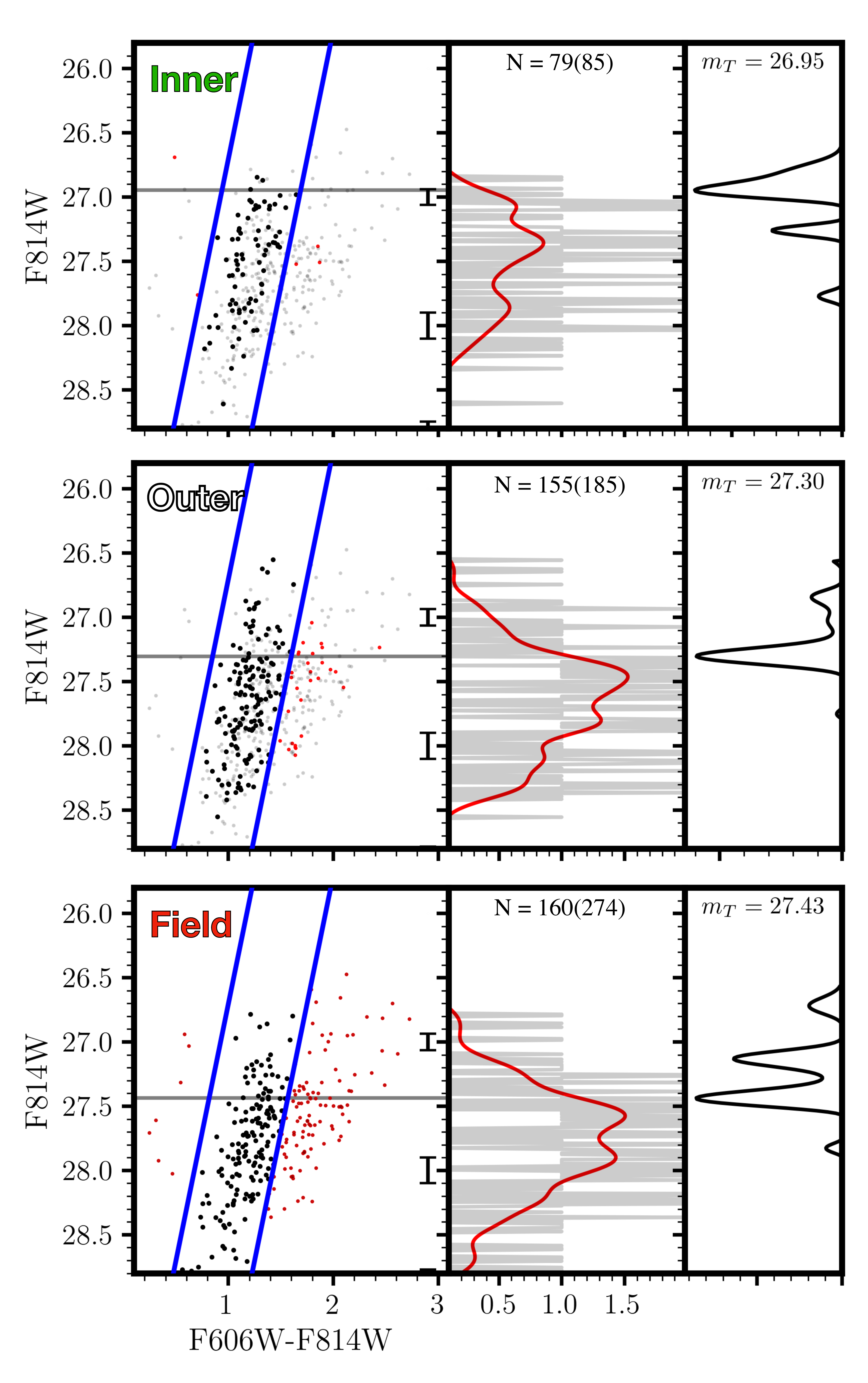}
    \caption{CMDs, LFs, and edge detections for the Inner, Outer, and Field regions (defined in \autoref{fig:dwarf1}). The color-selection derived from the Inner CMD is shown by blue parallel lines, and the sources excluded by it are plotted as red points. The number of sources after (before) the color selection are shown. Plotted as gray points in the Inner and Outer CMDs are ``Field'' (NGC~1404 main body) sources.
    }
    \label{fig:dwarf2}
\end{figure}

In \autoref{fig:dwarf2}, we present TRGB detection plots for each region. In each plot we demonstrate the effects of a narrow color selection with slope$= -4$~mag/mag, and $ 0.80 < $ (F606W$-$F814W) $ < 1.55 $~mag at $\mathrm{F814W}=27.3$~mag. 
As gray points we plot the sources belonging to the Field region and in red are those sources that were excluded by the color-magnitude selection.
This narrow color cut was determined from the Inner region's CMD, which should preferentially select for RGB stars that belong to FCC~B1281 and not NGC~1404. Indeed, the effectiveness of the color selection is confirmed by the decreasing number of sources contained within it for each progressively more extended region, with 93\%, 84\%, and 58\% for Inner, Outer, and Field, respectively. From comparing the locations of the with those in the Field region by way of the color selection, it becomes clear that the two populations are distinct. This can be considered a very coarse-grain implementation of filter-matching the morphology of the Inner CMD to that of the other two regions.

In \autoref{fig:dwarf2}, we see that the Inner region's stellar population is dominated by bright sources and the edge response peaks near F814W$=27$~mag. We suspect this brighter ``Tip'' could be due to a combination of crowding effects and potentially an increase in the AGB-to-RGB ratio in this Inner region, causing the edge detector to trigger off of a false Tip. For that reason, we do not consider this region further.

In the Outer CMD, there are an increased number of contaminant sources from NGC~1404 visible, but this region is still dominated by the same population associated with the Inner Region (as indicated by the color selection). The edge response is unambiguous and single-peaked, landing at F814W$=27.30$~mag.

The ``Field'' (NGC~1404) CMD definitively shows that the Inner and Outer regions contain a stellar population that is distinct from NGC~1404 because the locus of the ``Field'' TRGB lies outside of the color selection. This has a dramatic effect on the edge detection, where the kernel can be seen triggering off of the depleted, blue edge of the TRGB stars, leading to a faint edge detection at F814W$=27.43$~mag. Reassuringly, without the color-selection (i.e., inclusion of the red points) we measured a strongly peaked TRGB magnitude at F814W$=27.32$~mag, which exactly agrees with the value we determined from the full NGC~1404 catalog.

The fact that the color selection degraded the quality of the Field region's edge detection, but sharpened the edge measured in the Outer region, provides further evidence that the stellar populations measured in these two regions are distinct. As a result, we can conclude that the TRGB magnitude determined to the Outer region \mtsubscript$=27.30$~mag is indeed an independent distance measured via RGB stars belonging to FCC~B1281, and not NGC~1404. Based on the findings of \citet{madore_1995} on the determination of a TRGB magnitude in numbers-limited samples, we attribute a systematic error $\sigma_{sys} \sim 0.1$~mag to our TRGB distance to FCC~B1281.

We have determined a direct TRGB distance to FCC~B1281 that is within 0.02 mag of our measured distance to NGC~1404. Considering the on-sky proximity ($<\!30$~kpc) of FCC~B1281 to the center of NGC~1404 and their identical TRGB distances, we conclude that this dwarf satellite is very likely bound to NGC~1404. At the same time, the agreement in line-of-sight distances demonstrates the self-consistency of our TRGB distance measured to NGC~1404.

\section{Re-calibrating the CSP Distances to NGC~5643} \label{app:snia}
On its own, the SN~Ia distance scale is only determined relatively, between SNe, with a floating zero-point. This is acceptable for measurements of dark energy because the absolute calibration (i.e., dependence on the Hubble Constant) divides out. However, to measure absolute distances, one must adopt a calibration provided by external TRGB (or, e.g., Cepheid/SBF) distances measured to SN host galaxies. For that reason, any eventual measurement of the Hubble Constant will be perfectly degenerate with this calibration step (or rung) in the cosmic distance ladder.

For our purposes, we can take advantage of this degeneracy in order to bring the pair of SN~Ia distances to NGC~5643 measured by B20 onto the CCHP SN~Ia calibration.
B20 found from a multi-wavelength light curve fit $\mu_{2013aa}^{72} = 30.47 \pm 0.08$ mag and $\mu_{2017cbv}^{72} = 30.46 \pm 0.08$ mag, where the super-script denotes the value of their adopted $H_0$ value. As mentioned, there is a perfect degeneracy between $H_0$ and the absolute calibration of each rung in the distance ladder, so we need only calculate the shift in one value in order to simultaneously determine the shift in the other. The results of this recalibration are shown in \autoref{tab:snia}.
\begin{deluxetable}{lllll}
\tabletypesize{\normalsize} 
\tablewidth{0pt} 
\tablecaption{NGC~5643 SN~Ia Recalibration \label{tab:snia}} 
\tablehead{ 
\colhead{Name} &
\colhead{$\mu_{SN}$} &
\colhead{$H_0$} &
\colhead{Notes}
}
\startdata 
2013aa  & 30.47  & 72 & (a) \\
        & $30.54$ & 69.6 & (b) \\
2017cbv & 30.46  & 72 & (a)\\
        & $30.53$ & 69.6 & (b)
\enddata 
\tablecomments{(a) Distances from \citet{burns_2020}, $H_0 = 72$; (b) Rescaled from \citet{burns_2020} using $H_0 = 69.6$ \citep{freedman_2020}.}
\end{deluxetable}

\section{The Relative Distance between NGC~1404 and NGC~1316} \label{sect:1404_appendix}
\subsection{Computation of Relative Distances}

\begin{figure}
    \centering
    \includegraphics[width=0.9\columnwidth]{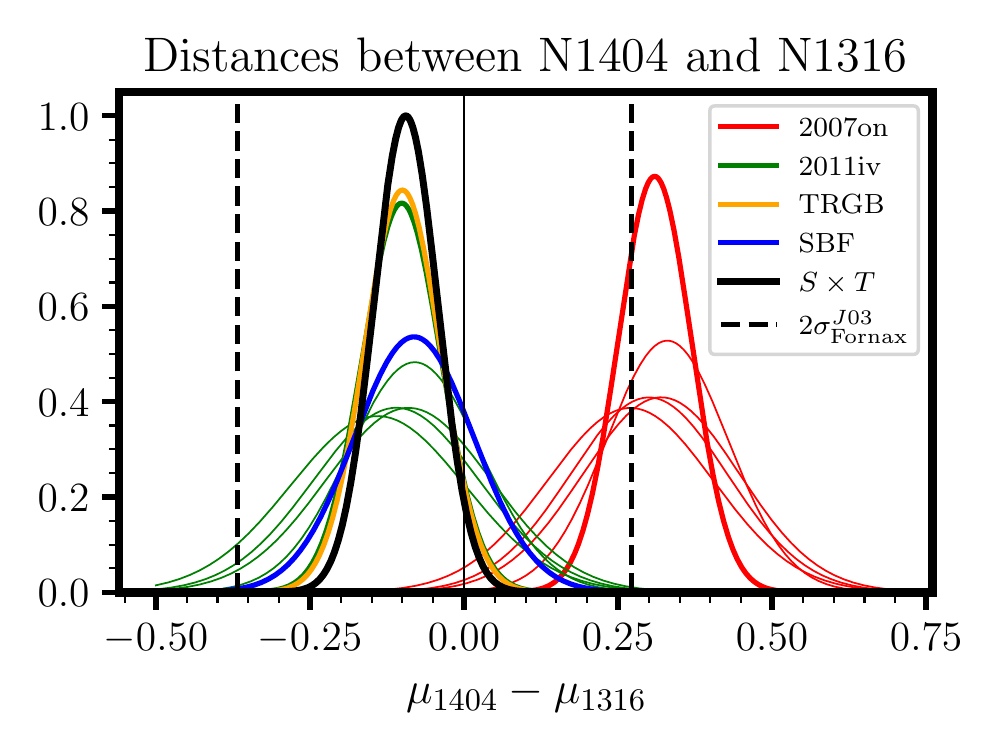}
    \caption{Relative distance $\Delta \mu\textit{} (\sigma)$ distributions determined with the TRGB (solid orange), SBF (solid blue), and SN~Ia (solid green and red) methods. The central value of each thick curve corresponds to the size of its respective arrow in \autoref{fig:dist_comp}. Plotted as the black curve is the product of the SBF and TRGB relative distances. Each of the plotted 2007on (solid red) and 2011iv (solid green) distributions is determined by subtracting the product of four separate relative distances (thin curves) calculated to each SN in NGC~1316. All distributions are re-normalized such that the peak value across them is unity.
    Ignoring the anomalously faint SN~2007on, the combined TRGB, SN, and SBF constraint on the relative line-of-sight distance between NGC~1404 and NGC~1316 is $\Delta \mu = -0.095 \pm 0.041$ mag.
    As a pair of vertical dashed lines we indicate twice the back-to-front depth of the Fornax Cluster \citep{jerjen_2003}. The midpoint of the two lines (i.e., center of the Fornax Cluster) is placed at half the average relative distance determined in this figure.
    }
    \label{fig:relative_comp}
\end{figure}

First, to calculate the relative distance for the TRGB method, we subtract from the NGC~1404 distance determined in this study the TRGB distance to NGC~1316 presented in \citet{freedman_2019}. This gives a value of $\Delta \mu_{TRGB} = \mu^{1404}_{TRGB} - \mu^{1316}_{TRGB} = \FOURTEENdiffTHIRTEENround \pm \FOURTEENdiffTHIRTEENerrround $ mag. In computing the uncertainty, we have removed any contributions from the ACS absolute photometric calibration and the TRGB zero-point. The absolute distances (blue star and black triangle) and the relative distance (green arrow) were plotted at the bottom of \autoref{fig:diff_mod}. In \autoref{fig:relative_comp}, the relative distance $\Delta \mu_{TRGB} $ is plotted as the orange curve. 

Then, to compute the SBF relative distance, we adopt the values measured in \citet{Blakeslee_2009}. We choose this set of distances because they were measured homogeneously as part of the same study, further minimizing potential systematic errors due to author methodology or calibrating sample. They determined an SBF distance to NGC~1404 $\mu_{SBF} = 31.526 \pm 0.072 $ mag and for NGC~1316 $\mu_{SBF} =  31.607 \pm 0.065$ mag, such that $\Delta \mu_{SBF} = \mu^{1404}_{SBF} - \mu^{1316}_{SBF} = -0.081 \pm 0.097 $ mag. The absolute distances (black square and black triangle) and the relative distance (green arrow) were plotted at the top of \autoref{fig:diff_mod}. In \autoref{fig:relative_comp}, $\Delta \mu_{SBF} $ is plotted as the blue curve.

Finally, for the SN~Ia relative distance, we adopt the distances measured by \citet{burns_2020}. We again adopt the CSP SNe~Ia distances to remain consistent with the \citet{freedman_2019} cosmology sample. They fit the four SNe in NGC~1316 using the SNooPY package and, assuming an $H_0 = 72$ km/s/Mpc, found distance moduli 31.27, 31.32, 31.29, 31.26 mag, with errors 0.09, 0.10, 0.09, 0.04 mag, for SNe 1980N, 1981D, 2006dd, 2006mr, respectively. These distances were plotted in the middle of \autoref{fig:diff_mod} as black triangles. Similarly for NGC~1404, the same CSP SNooPY fits yield moduli $ \mu = 31.59 \pm 0.09 $~mag and $ 31.18 \pm 0.10 $~mag, for SN 2007on and 2011iv, respectively. Notably, the two SNe are discrepant in their measured distances by 0.4 mag \emph{after} correction for their (nearly identical) decline rates \citep{gall_2018}. These four distances are plotted as black squares in the middle row of \autoref{fig:diff_mod}.

\begin{figure}
    \centering
    \includegraphics[width=\columnwidth]{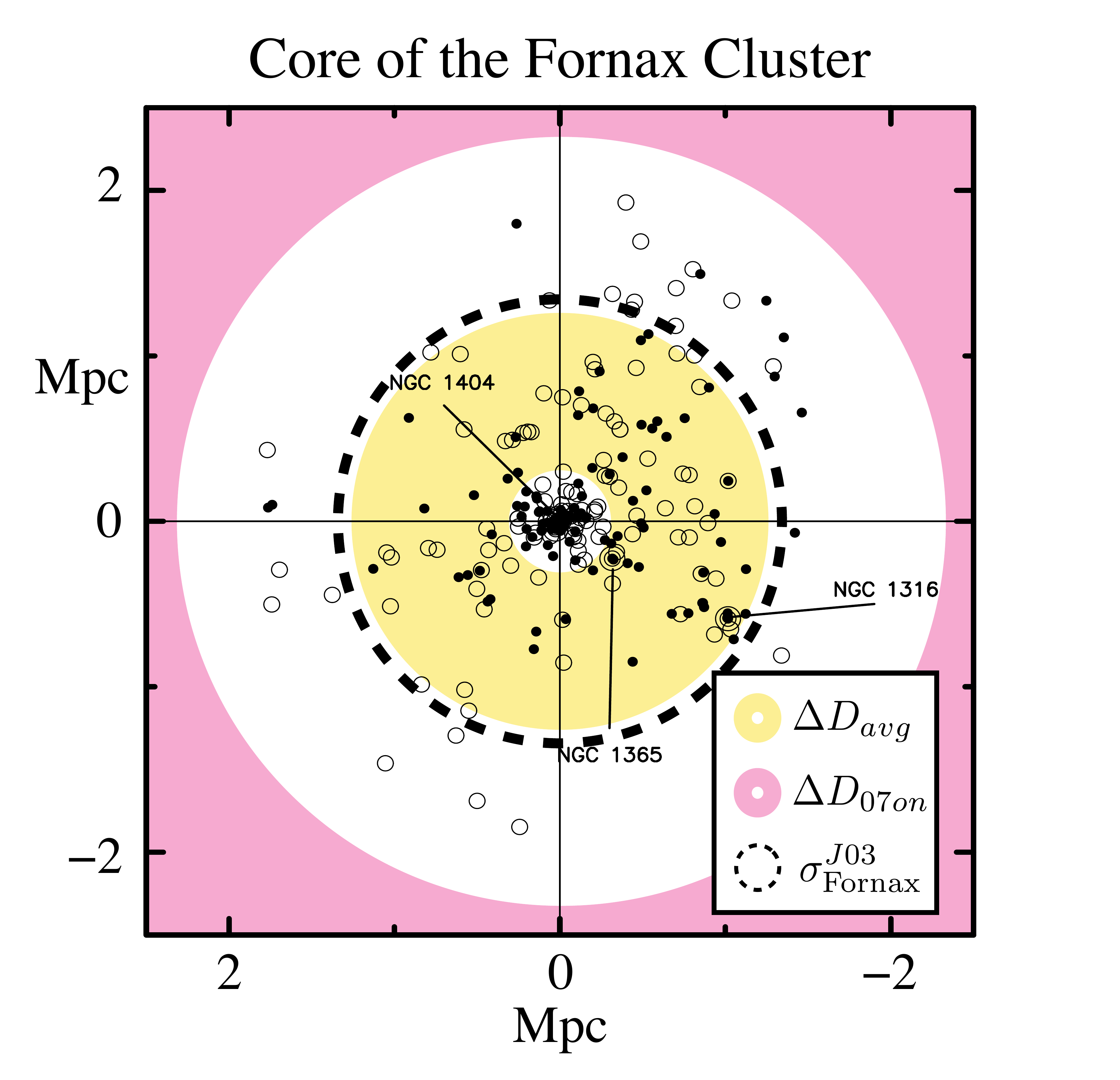}
    \caption{Radial velocity confirmed members located in the core of the Fornax Cluster, from the NED Environment Search tool (\url{https://ned.ipac.caltech.edu/forms/denv.html}). The positions on the sky for each galaxy are transformed to physical units relative to NGC~1399, assuming a line-of-sight distance $D = \distGALTWOround$~Mpc. Open (closed) circles denote negative (positive) radial velocities relative to the adopted mean velocity of the Cluster. Labeled are three galaxies to which the CCHP has measured a TRGB distance. The yellow, shaded region represents the 1-sigma interval for the line-of-sight quantity $\Delta \mu^{S \times T}_{1404-1316}$ in physical units $\Delta D_{avg}$ after projection onto the 2D sky. Similarly, the quantity $\Delta \mu^{\mathrm{07on}}_{1404-1316}$ is transformed to $\Delta D_{\mathrm{07on}}$ and plotted as a magenta shaded region, showing the expected location of the SNe in NGC~1316 relative to SN~2007on in NGC~1404. The dashed circle represents the depth of the Fornax Cluster \citep{jerjen_2003}. The magenta-shaded region extends well beyond the boundaries of the figure, and thus the physical extent of the Fornax Cluster.
    }
    \label{fig:fornax_map}
\end{figure}

For each SN we compute the relative distance between each combination, i.e.,
\begin{align}
\Delta \mu_{ij} = \mu_{i} - \mu_{j}, ~ \forall i & \in \{ \mathrm{2007on}, \mathrm{2011iv}\}, \\
                          j & \in \{ \mathrm{1980N}, \mathrm{1981D}, \mathrm{2006dd}, \mathrm{2006mr} \} \nonumber
\end{align}
where each uncertainty $\sigma_{ij}$ is computed by adding the individual distance uncertainties in quadrature. In \autoref{fig:relative_comp} each $\Delta \mu(\sigma)_{ij}$ is plotted as a thin green, or thin red curve. Then, because the SN distances to NGC~1316 are in good agreement, we compute the product distributions
\begin{equation}
\Delta \mu(\sigma)_{i} = \prod_j \Delta \mu(\sigma)_{ij}
\end{equation}
In the middle row of \autoref{fig:diff_mod} we plot the peak value of $\Delta \mu(\sigma)_{\mathrm{2007on}}$ as a red arrow and of $\Delta \mu(\sigma)_{\mathrm{2011iv}}$ as a green arrow. In \autoref{fig:relative_comp}, each $\Delta \mu(\sigma)_{i}$ is plotted as a thick curve, where the color matches the corresponding arrow plotted in \autoref{fig:diff_mod}. Because both of the $\Delta \mu(\sigma)_{i}$ quantities result from a subtraction with the same four SNe in NGC~1316 they are, of course, correlated with one another.

Combining the TRGB and SBF relative distances we determine $\Delta \mu_{S \times T} = -0.094 \pm 0.057$~mag. This is plotted in \autoref{fig:relative_comp} as a black curve. Taking the combined SBF+TRGB relative distance, we can immediately compute the sigma significance at which SN~2007on is anomalously faint with respect to the SNe in NGC~1316,
$$\frac{ \Delta \mu_{S\times T} - \Delta \mu_{07on}} {\sqrt{\sigma_{S\times T}^2 + \sigma_{07on}^2}} $$ 
which gives a value of 4.9 sigma. We note that this calculation does not utilize SN~2011iv in any way. 

As an additional external constraint, \cite{jerjen_2003} used their SBF distance measurements to the dwarf elliptical population in the Fornax Cluster and estimated its depth to be ($ \sigma_{\mathrm{Fornax}}^{J03} = 0.16^{+0.06}_{-0.09} $ mag). The relative distance we have calculated between NGC~1316 and NGC~1404 falls well within this range, as expected for the more massive galaxies in the Cluster. However, the difference in inferred distance between SN~2007on and SN~2011iv falls well outside of this range, providing further evidence that 2007on is not reliable as a distance indicator. In \autoref{fig:relative_comp}, vertical dashed lines are shown at twice the size of this interval.

In \autoref{fig:fornax_map} we present an alternative way of viewing the relative distances computed in this section, as well as visualizing the combined evidence for why the fainter SN~2007on may not suitable for use as a distance indicator. Plotted are the on-sky positions of 259 members of the Fornax Cluster, where membership was determined based on a radial velocity cut $v_r = 1379 \pm 500$km~s$^{-1}$. The physical scale is determined by assuming a distance of \distGALTWOround~Mpc. We have labeled NGC~1404, NGC~1365, and NGC~1316; each is a galaxy to which the CCHP has measured a TRGB distance. Plotted as a black dashed line is again the estimated depth of the Fornax Cluster $\sigma_{\mathrm{Fornax}}^{J03}$, projected onto the 2D sky. The $1\sigma$ interval of the combined SBF-TRGB relative distance is ``de-projected'' onto the sky as a yellow disk. The distance between the four SNe in NGC~1316 and SN~2007on in NGC~1404 is also represented on the sky as a pink disk that extends beyond the bounds of the plot.

The high significance of this determination that 2007on is subluminous could suggest that (1) 2007on is located anomalously in the background of the Fornax Cluster, and not near NGC~1404 itself (not likely), (2) the transitional types are drawn from a population of SNe that exhibit a higher dispersion in their luminosity-width relation, and that current light curve fits are not capturing the full uncertainties, or (3) 2007on resulted from a different progenitor event than that of ``normal'' Ia's \citep[see, e.g.,][]{Wygoda_2019}. This would suggest that any future cosmological analyses that use transitional type SNe should consider targeted statistical tests (e.g., jackknifing, etc.) to quantify how the ex-/in-clusion of these objects might affect the results.

\bibliographystyle{aasjournal}
\bibliography{bib_trgb,bib_cchp,bib_cosmo,bib_dists,bib_gals,bib_main,bib_phot,bib_snia}{}
\end{document}